\documentclass[a4paper,fleqn,usenatbib]{mnras}

\usepackage{hyperref}  % Hyperlinks 
\hypersetup{colorlinks=true,linkcolor=blue,citecolor=blue,filecolor=blue,urlcolor=blue}

\usepackage{newtxtext,newtxmath,ifpdf}
\usepackage[T1]{fontenc}
\usepackage{ae,aecompl}

\usepackage{graphicx}	% Including figure files
\usepackage{amsmath}	% Advanced maths commands
\usepackage{amssymb}	% Extra maths symbols

\usepackage{graphicx,lscape,amssymb,amsmath,url,placeins,mathrsfs,float,bm}

\def\aap{A\hbox{\rm \&}A} 
  \def\aj{AJ} \def\ao{Appl.\ Opt.}
  \def\apj{ApJ} \def\apjl{ApJ}
 \def\apjs{ApJS} 
\def\araa{ARA\hbox{\rm \&}A}

\def\mnras{MNRAS} \def\nat{Nat}

\def\physrep{{Phys.~Rep.}}   % Physics Reports

\title[AS2UDS: Dust attenuation in ALMA-detected LBGs]{An ALMA survey of the SCUBA-2 Cosmology Legacy Survey UKIDSS/UDS field: Dust attenuation in high-redshift Lyman break Galaxies}

\author[M.\ Koprowski et al.]
{M.\,P.\ Koprowski$^{1,2}$\thanks{E-mail: mkoprowski@fizyka.umk.pl},
K.\,E.\,K.~Coppin$^{1}$,
J.\,E.\ Geach$^{1}$,
U.\ Dudzevi\v{c}i\={u}t\.{e}$^3$,
Ian Smail$^3$,
\newauthor
O.\ Almaini$^4$,
Fangxia An$^3$,
A.\,W.\ Blain$^5$,
S.\,C.\ Chapman$^6$,
Chian-Chou Chen$^{7,8}$,
\newauthor
C.\,J.\ Conselice$^4$,
J.\,S.\ Dunlop$^9$,
D.\ Farrah$^{10,11}$,
B.\ Gullberg$^3$,
W.\ Hartley$^{12,13}$,
\newauthor
R.\,J.\ Ivison$^{7,9}$,
A.\ Karska$^2$,
D.\ Maltby$^4$,
M.\,J.\ Micha{\l}owski$^{14}$,
A.\ Pope$^{15}$
S.\ Salim$^{16}$,
\newauthor
D.\ Scott$^{17}$,
C.\,J.\ Simpson$^{18}$,
J.\,M.\ Simpson$^{19}$,
A.\,M.\ Swinbank$^3$,
A.\,P.\ Thomson$^{20}$,
\newauthor
J.\,L.\ Wardlow$^{21}$,
P.\,P.\ van der Werf$^{22}$,
K.\,E.\ Whitaker$^{23}$
\\
$^1$Centre for Astrophysics Research, School of Physics, Astronomy and Mathematics, University of Hertfordshire, College Lane, Hatfield AL10 9AB, UK\\
$^2$Institute of Astronomy, Faculty of Physics, Astronomy and Informatics, Nicolaus Copernicus University, Grudziadzka 5, 87-100 Torun, Poland\\
$^3$Centre for Extragalactic Astronomy, Department of Physics, Durham University, Durham, DH1 3LE, UK\\
$^4$School of Physics and Astronomy, University of Nottingham, University Park, Nottingham NG7 2RD, UK\\
$^5$Physics \& Astronomy, University of Leicester, 1 University Road, Leicester LE1 7RH, UK\\
$^6$Department of Physics and Atmospheric Science, Dalhousie University, Halifax, NS B3H 4R2, Canada\\
$^7$European Southern Observatory, Karl Schwarzschild Strasse 2, Garching, Germany\\
$^8$Academia Sinica Institute of Astronomy and Astrophysics (ASIAA), No. 1, Section 4, Roosevelt Rd., Taipei 10617, Taiwan\\
$^9$Institute for Astronomy, University of Edinburgh, Royal Observatory, Edinburgh EH9 3HJ, UK\\
$^{10}$Department of Physics and Astronomy, University of Hawaii, 2505 Correa Road, Honolulu, HI 96822, USA\\
$^{11}$Institute for Astronomy, 2680 Woodlawn Drive, University of Hawaii, Honolulu, HI 96822, USA\\
$^{12}$Department of Physics and Astronomy, University College London, London, WC1E 6BT, UK\\
$^{13}$Department of Physics, ETH Zurich, Wolfgang-PauliStrasse 16, CH-8093 Zurich, Switzerland\\
$^{14}$Astronomical Observatory Institute, Faculty of Physics, Adam Mickiewicz University, ul.~S{\l}oneczna 36, 60-286 Pozna{\'n}, Poland\\
$^{15}$Department of Astronomy, University of Massachusetts, Amherst, MA 01003, USA\\
$^{16}$Department of Astronomy, Indiana University, Bloomington, IN 47404, USA\\
$^{17}$Department of Physics and Astronomy, 6224 Agricultural Road, University of British Columbia, Vancouver V6T 1Z1, Canada\\
$^{18}$Gemini Observatory, Northern Operations Center, 670 N. A’ohuku Place, Hilo, HI 96720, USA\\
$^{19}$Academia Sinica Institute of Astronomy and Astrophysics, No. 1, Sec. 4, Roosevelt Rd., Taipei 10617, Taiwan\\
$^{20}$The University of Manchester, Oxford Road, Manchester, M13 9PL, UK\\
$^{21}$Department of Physics, Lancaster University, Lancaster, LA1 4YB, UK\\
$^{22}$Leiden Observatory, Leiden University, P.O. Box 9513, NL-2300 RA Leiden, The Netherlands\\
$^{23}$Department of Astronomy, University of Massachusetts, Amherst, MA 01003, USA
}

\date{Accepted XXX. Received YYY; in original form ZZZ}

\pubyear{2020}
\begin{document}
\label{firstpage}
\pagerange{\pageref{firstpage}--\pageref{lastpage}}
\maketitle

\begin{abstract}

We analyse 870$\mu$m Atacama Large Millimetre Array (ALMA) dust continuum detections of 41 canonically-selected $z\simeq 3$ Lyman-break galaxies (LBGs), as well as 209 ALMA-undetected LBGs, in follow-up of SCUBA-2 mapping of the UKIDSS Ultra Deep Survey (UDS) field. %Spectral energy distribution (SED) fitting across UV to far-infrared (FIR) wavelengths was performed, in order to quantify various physical properties and compare to more typical FIR-faint LBGs selected in the footprints of the same ALMA maps. 
We find that our ALMA-bright LBGs lie significantly off the locally calibrated IRX-beta relation and tend to have relatively bluer rest-frame UV slopes (as parametrised by $\beta$), given their high values of the `infrared excess' (IRX $\equiv L_{\rm IR}/L_{\rm UV}$), relative to the average `local' IRX-$\beta$ relation. We attribute this finding in part to the young ages of the underlying stellar populations but we find that the main reason behind the unusually blue UV slopes are the relatively shallow slopes of the corresponding dust attenuation curves. We show that, when stellar masses, $M_\ast$, are being established via SED fitting, it is absolutely crucial to allow the attenuation curves to vary (rather than fixing it on Calzetti-like law), where we find that the inappropriate curves may underestimate the resulting stellar masses by a factor of $\simeq$2-3$\times$ on average. In addition, we find these LBGs to have relatively high specific star-formation rates (sSFRs), dominated by the dust component, as quantified via the fraction of obscured star formation ($f_{\rm obs}\equiv {\rm SFR_{\rm IR}/{\rm SFR}_{\rm UV+IR}}$)%, placing them well above the $z<2.5$ $f_{\rm obs}$-$M_\ast$ relation
. We conclude that the ALMA-bright LBGs are, by selection, massive galaxies undergoing a burst of a star formation (large sSFRs, driven, for example, by secular or merger processes), with a likely geometrical disconnection of the dust and stars% (enabling these to be simultaneously selected both as LBGs and FIR-bright sources in the first place)
, responsible for producing shallow dust attenuation curves.

\end{abstract}

% \begin{keywords}
% dust, extinction -- galaxies: evolution, high-redshift, star formation, ISM -- cosmology: observations
% \end{keywords}

%%%%%%%%%%%%%%%%%%%%%%%%%%%%%%%%%%%%%%%%%%%%%%%%%%

%%%%%%%%%%%%%%%%% BODY OF PAPER %%%%%%%%%%%%%%%%%%

%
%
%
\section{Introduction}
\label{sec:intro}

One of the most fundamental goals of galaxy evolution studies is the robust measurement of the cosmic history of star formation \citep{Madau_2014}, the most direct tracer of which is the rest-frame UV light, dominated by the emission from massive stars \citep{Kennicutt_2012}. The UV luminosity, with the aid of the assumed stellar initial mass function (IMF), is therefore the most widely used observable in determining the corresponding star formation, particularly at high redshifts (e.g.\ \citealt{Bouwens_2011, Dunlop_2012, Ellis_2013, Oesch_2014, Oesch_2015, Bouwens_2015, McLeod_2015, McLeod_2016}). However, the use of the UV luminosity as a tracer of the star-formation rate (SFR) is hampered by the absorption of the UV light by the dust present in the interstellar medium (ISM). The energy absorbed by dust is re-emitted in the infrared (IR) and, therefore, one ideally requires the IR observations to complement the UV data, in order to paint complete picture of star formation. 

At $z>2$, most studies of the evolution of the star-formation rate density (SFRD) are based on samples of Lyman-break galaxies (LBGs), due in part because of the efficiency of their selection technique in deep broad-band imaging surveys. As a result, LBGs have been extensively studied and well characterised over the past two decades. With stellar masses of $\sim 10^{9-11}\,{\rm M_\odot}$ and star-formation rates (SFRs) of $\sim$\,$10$--$100$\,${\rm M_\odot\,yr^{-1}}$ (e.g.\ \citealt{Madau_1996, Reddy_2006, Stark_2009, Oteo_2013}), LBGs have been suggested to be responsible for forming a substantial fraction of the stellar mass in massive local galaxies ($L\geq {\rm L_\ast}$; e.g.\ \citealt{Somerville_2001, Baugh_2005}), with those with the highest SFRs ($>100\,{\rm M_\odot\, yr^{-1}}$) being potentially the progenitors of present-day ellipticals (e.g.\ \citealt{Verma_2007, Reddy_2009, Stark_2009}).  However, the influence of dust extinction is expected to have a significant impact on the derived estimates of the star-formation rates of individual sources, as well as the completeness of these UV-selected studies. Hence, it is important to try to constrain the far-infrared (FIR) luminosities of LBGs to derive more robust estimates of their SFRs.  Unfortunately, due to the relatively modest sensitivity of the single-dish far-infrared and sub-/millimetre facilities and the rather small fields of view of sub/-millimetre interferometers, the vast majority of LBGs lack strong constraints on their rest-frame far-infrared or sub-millimetre emission and hence alternative methods have to be used in order to account for the dust-absorbed UV light (e.g. \citealt{Chapman_2000, Chapman_2002b, Webb_2003, Coppin_2015}).

The most common approach to estimating the far-infrared luminosity of UV-selected galaxies is to use the observed correlation between the so-called infrared excess, ${\rm IRX}\equiv L_{\rm IR}/L_{\rm UV}$, and the measured UV slope, $\beta$, (e.g.\ \citealt{Meurer_1999, Overzier_2011, Takeuchi_2012}).  Dust obscuration becomes progressively less important towards near-IR wavelengths (dust extinction is about ten times lower at 2.2\,${\rm \mu m}$ than at 0.55\,${\rm \mu m}$; \citealt{Calzetti_1997}), so that the slope of the rest-frame UV stellar emission spectrum, $\beta$, can be used as an indication of the amount of dust attenuation (with $F_\lambda\propto \lambda^\beta$). Although it should be stressed that this estimate only applies to the {\it detected} UV continuum emission (if emission is too highly obscured then it will be completely missed) and is weighted towards the least extincted parts of the source. 
It has been found that IRX correlates tightly with the observed UV slope for local UV-selected starburst galaxies \citep{Calzetti_1997, Meurer_1999}, as well as similarly UV-bright high-redshift sources (e.g.\ \citealt{Seibert_2002, Reddy_2008, Reddy_2012b, Pannella_2009, Heinis_2013, Bourne_2017, McLure_2018}). The local IRX-$\beta$ relation is, therefore, often used in order to attempt to correct for the dust absorption in high redshift ($z>2$) galaxies, when no IR constraints are available (e.g.\ \citealt{Treyer_2007, Gonzalez_2010, Gonzalez_2014, Bothwell_2011, Finkelstein_2012, Smit_2012, Smit_2016}).

More recent work has revealed that the local IRX-$\beta$ relation is not as tight as initially claimed, with significant scatter between individual galaxies (e.g.\ \citealt{Kong_2004}). It has been proposed that the inconsistencies may be explained with the introduction of a so-called `third parameter', examples of which include different ages of the underlying stellar populations (e.g.\ \citealt{Kong_2004, Burgarella_2005, Boquien_2009, Grasha_2013}), variations in the shapes of the intrinsic (without dust attenuation) UV slopes (e.g.\ \citealt{Boquien_2012}), and differences in dust types (e.g.\ \citealt{Thilker_2007}). Although the most extreme outliers may simply reflect a fundamental mismatch between the emission regions traced by the less-obscured UV and more-obscured FIR emission \citep{Goldader_2002, Trentham_1999}.  More recently, it has been shown that the scatter in the IRX-$\beta$ relation may be driven by the diversity in the corresponding attenuation curves \citep{Salmon_2016, Salim_2019}.  

Similarly, at high redshifts, {\it Herschel} and ALMA observations have shown that the most luminous FIR galaxies, when detected in the restframe UV, tend to have bluer UV slopes at fixed IRX (e.g.\ \citealt{Reddy_2010, Penner_2012, Oteo_2013, Casey_2014b, Watson_2015}). \citet{Casey_2014b} relates the deviation bluewards from the local IRX-$\beta$ relation to the IR luminosity, where they found that more luminous IR galaxies exhibit bluer UV colours. At the same time, it has been claimed that some high-redshift sources lie below the local relation \citep{Reddy_2010, Whitaker_2012, Capak_2015, Koprowski_2016, Bouwens_2016b, Pope_2017}, which has been suggested to be indicative that stellar light in these sources is affected by a different (e.g.\ SMC-like) dust extinction law.

Various theoretical efforts have attempted to explain the apparent deviations from the local IRX-$\beta$ relation \citep{Ferrara_2017, Safarzadeh_2017, Popping_2017, Narayanan_2018}. It was proposed that the most probable reason for the blue, dusty star-forming galaxies lying above the local relation, is the irregular relative covering of the dust and stars, where the least dust-attenuated regions produce blue UV colours, while the dusty parts give rise to the high values of the IRX. On the other hand, the galaxies with redder UV colours may be characteristic of older stellar populations, or affected by different dust extinction laws. Another possible explanation could be that the assumed dust temperatures for these sources are too low, leading to an underestimation of the IR luminosities and thus IRX.

In this work, we were able to precisely investigate the dust attenuation properties of high-redshift Lyman-break galaxies, as quantified via the IRX-$\beta$ relation, thanks to the availability of the FIR data. Matching $\sim$8000 3$\lesssim$$z$$\lesssim$5 LBGs with a sample of 708 submm-luminous sources, detected as a part of the ALMA survey of sub-millimeter galaxies from the SCUBA-2 Cosmology Legacy Survey (S2CLS, \citealt{Geach_2017}) map of the UKIDSS Ultra Deep Survey (UDS) field \citep{Stach_2019}, we identified an exquisite sub-sample of 41 ALMA-bright Lyman-break galaxies, for which we have the high-resolution dust continuum detections. Using rest-frame UV-to-FIR SED fitting, we are able to quantify the physical properties of the ALMA-detected LBGs, and we compare these to the more typical, FIR-faint high-redshift LBGs (using 209 LBGs in the ALMA coverage which are undetected in those maps).

In Section\,\ref{sec:data} we describe the data used in our analysis. Section\,\ref{sec:analysis} describes how the SED fitting was performed. It includes a quantitative description of the method used for varying the assumed shapes of the attenuation curves (Section\,\ref{sec:sed}), a summary of the derived physical properties, including stellar masses and UV \& IR luminosities (Section\,\ref{sec:props}), as well as the methodology for measuring $\beta$ and IRX (Section\,\ref{sec:irx1}). Section\,\ref{sec:disc} provides a discussion of the results, and our main conclusions are summarised in Section\,\ref{sec:summ}.
%description of the physics behind the unusually blue UV slopes in the ALMA-detected LBGs compared with the average properties of the FIR-faint LBGs. We discuss the relationship between the IRX and the stellar mass in Section\,\ref{sec:irxm}. The fraction of obscured star formation hidden in the dust component for our sample is briefly investigated in Section\,\ref{sec:fobs}, while the impact of the assumed attenuation law on the inferred stellar masses is discussed in Section\,\ref{sec:mass}. We summarise in Section\,\ref{sec:summ}.

Throughout the paper we use the \citet{Chabrier_2003} stellar IMF with assumed flat cosmology with $\Omega_{\rm m} = 0.3$, $\Omega_\Lambda = 0.7$ and H$_0 = 70 $\,km\,s$^{-1}$\,Mpc$^{-1}$. 

% We note that assuming the best-fit \citet{Planck_2016} cosmology yields $\simeq 2-2.5\%$ higher luminosity distances as well as $\simeq 4-5\%$ higher stellar masses and luminosities.

%
%
%
\section{Data}
\label{sec:data}

The starting point for our analysis is the  catalogue of
716 sub-millimetre sources detected in the 850-${\rm \mu m}$ map of the UDS field from the SCUBA-2 Cosmology Legacy Survey \citep{Geach_2017}.  These were subsequently followed up with ALMA 870\,${\rm \mu m}$ (Band-7) imaging  between November 2013 and May 2017, 
with an initial pilot study of 30 of the brightest SCUBA-2 sources in Cycle 1 by \cite{Simpson_2017},  see also \citet{Simpson_2015,Simpson_2015b}, and then subsequent observations to complete the full sample in Cycles 3, 4 and 5 by \citet{Stach_2019}, see also \citet{Stach_2018}. For the completed AS2UDS survey,
sources were identified  by \citet{Stach_2019} using cleaned, 0.5 arcsec FWHM-tapered continuum maps, with average depths of $\sigma_{870}=0.25, 0.34, 0.23 $ and  0.09 mJy beam$^{-1}$ (Cycles 1, 3, 4 and 5, respectively). \citet{Stach_2019} catalogued 708 sub-millimetre galaxies (SMGs) brighter than $S_{870}\gtrsim 1$ mJy (4.3$\sigma$, or a false-positive rate of 2\%) in the full sample. For more details of the calibration,  analysis and catalogue see \citet{Stach_2019}.

The detection limit of $S_{870}=1$ mJy at $z$$\simeq$3 corresponds to the infrared luminosity of $\sim$$10^{12}\,{\rm L_\odot}$. It is, therefore, comparable to the individual detections for SCUBA2 850\,${\rm \mu m}$ deep surveys (e.g. \citealt{Geach_2017}) and somewhat deeper than the {\it Herschel} studies at these redshifts (e.g., \citealt{Casey_2014}), but significantly shallower than the luminosities reached with stacking of the high-redshift LBGs in the FIR maps (e.g., \citealt{Coppin_2015, Koprowski_2018}), as well as some of the other LBGs individual ALMA detections (e.g., \citealt{Capak_2015, Koprowski_2016}).

Photometric redshifts and multi-wavelength properties for the SMGs in the AS2UDS sample are presented in \citet{Dudzeviciute_2019}, where the full SEDs across the rest-frame UV-radio photometry, were modelled using the high-redshift version of the {\sc magphys} code \citep*{Cunha_2008, Cunha_2015, Battisti_2019}. This analysis includes the sub-millimetre imaging from the {\it Herschel} \citep{Pilbratt_2010}, as provided by the public releases of the HerMES \citep{Oliver_2012} and PEP \citep{Lutz_2011} surveys undertaken with the SPIRE \citep{Griffin_2010} and PACS \citep{Poglitsch_2010} instruments. Due to the large beam sizes of the {\it Herschel}-SPIRE data, deblended maps were produced following the de-blending procedure detailed in \citet{Swinbank_2014}, with ALMA, MIPS 24\,${\rm \mu m}$ and radio catalogue sources used as priors for the locations of sources contributing to the PACS and SPIRE map flux.  See \citet{Dudzeviciute_2019} for more details of the photometry and {\sc magphys} analysis.

We matched the SMGs from the ALMA Band-7 $>4.3\,\sigma$ catalogue with the UDS LBG sample from \citet{Koprowski_2018}. We start by  updating the LBG selection using the new  UV/optical/near-IR DR11 UDS catalogue (from the DR8 used previously). The primary benefit is the improved depth of 0.7 magnitudes in $K$-band (Almaini et al.\ in prep.; Hartley et al.\ in prep.). We identify galaxies at $z$$\simeq$3 using UGR, or BVR, filters \citep{Steidel_1996}, extending to higher redshifts by simply shifting the colour space to longer wavelengths, as described by \citet{Ouchi_2004}, finding 8494 3$\lesssim$$z$$\lesssim$5 LBGs. Since the parent optical catalogue is selected at $K$-band ($K<25.3$), our resulting LBG sample is mass complete to a limit of ${\rm log}(M_\ast/{\rm M_\odot})\gtrsim9.5$.

In order to select the optimal search radius for the cross-matching (minimizing the false detection rate, whilst maximizing the number of associations), we performed a suite of Monte Carlo simulations in the following manner. In each of 1,000 realizations, an artificial LBG catalogue was constructed by shifting all the individual sources in the original catalogue in random directions over a distance of 10\,arcsec. The average number of random ALMA-LBG associations can then be found as a function of search radius and compared with the unshifted map cross-matching to yield a ratio of the false association rate. In this way, we found a false association rate of $\sim 3\%$ for a search radius of 0.7\, arcsec, rising to $\sim 10\%$ at 1.5\, arcsec and $\sim 30\%$ at 2.2\, arcsec. Out of 250 3$\lesssim$$z$$\lesssim$5 LBGs in the UDS field which fall within the primary beam coverage of the AS2UDS ALMA maps, 41 were found to have ALMA Band-7 detections ($\geq 4.3\sigma$) within a 0.7 arcsec search radius (corresponding to $\sim$\,5\,kpc at $z\sim 3$--$5$), where only one of our optical counterparts is statistically expected to be a false association. %This yields a 16$\pm$3\% detection fraction of LBGs with ALMA, although we note that these are not blank-field maps, since they are centred on biased locations in the sky already known to harbour massive FIR-bright galaxies, so the actual detection rate, considering the whole LBG population in the UDS field, as described in \citet{Koprowski_2018}, is $\simeq$3\%. 
We note that our analysis makes use of the remaining $\sim 200$ ALMA-undetected LBGs for which the 870-$\mu$m maps provide limits on their FIR emission.

\begin{figure*}
\begin{center}
\includegraphics[scale=0.77]{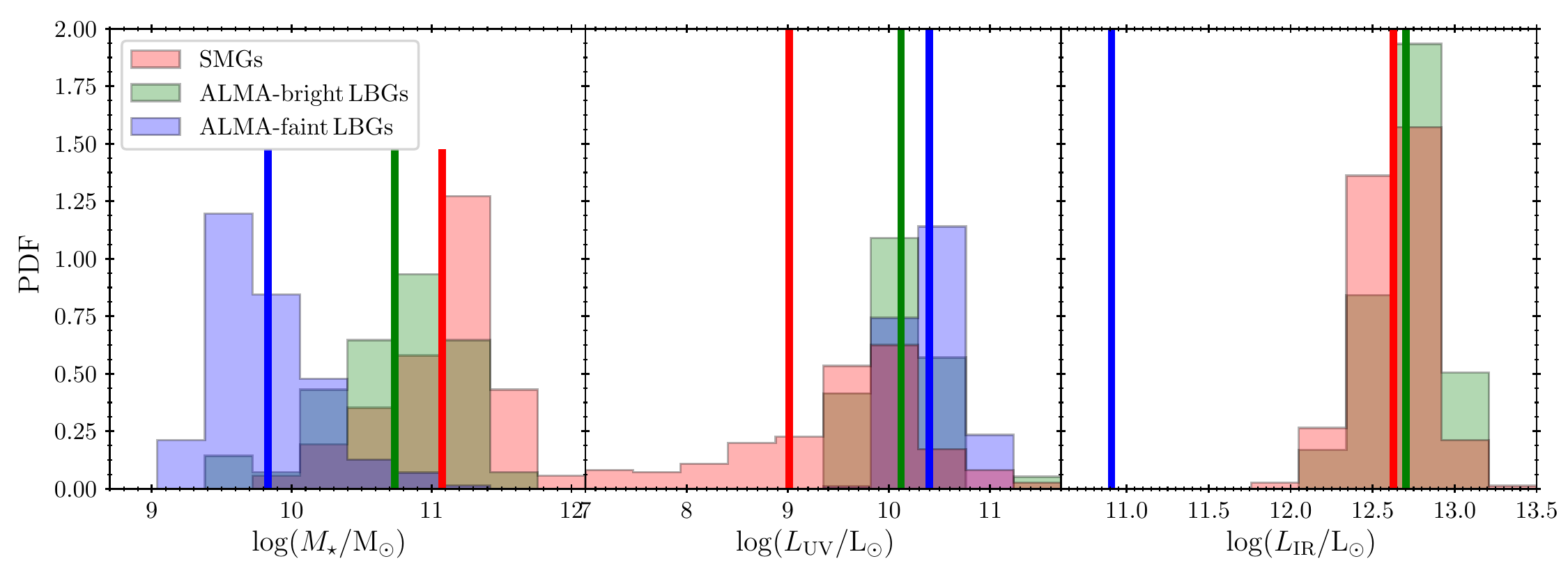}
\end{center}
\caption{Histograms of main physical properties for FIR-bright LBGs, their FIR-faint analogues and the ALMA-detected SMGs that were not selected as LBGs, with the vertical lines representing mean values. The ordinate represents the probability density function, with the area under the curve equal unity. The left panel shows stellar masses, derived from the SED fits to the available UV-FIR photometry, assuming varying-slope attenuation curves, while for the SMGs stellar masses were adopted from \citet{Dudzeviciute_2019}. The middle panel depicts the distribution of UV luminosities found from best-fit SEDs, while the right panel shows values of IR luminosities. The blue vertical line in the right panel represents the stack of LBGs lying within the ALMA stamps that were not individually detected in the FIR luminosity of $L_{\rm IR}=6.6^{+2.1}_{-1.6} \times 10^{10} {\rm L}_\odot$ (see Section\,\ref{sec:props}). It can be seen that the ALMA-bright LBGs tend to have much higher stellar masses and IR luminosities, yet also slightly lower UV luminosities, on average, compared with the ALMA-undetected LBGs, which together suggest that the ALMA-bright LBGs suffer from correspondingly higher rest-frame UV dust obscuration on average. When comparing to the SMGs that have not been classified as LBGs, ALMA-bright LBGs seem to sample the full SMG distribution, with no obvious biases towards the high/low $M_\ast$/$L_{\rm IR}$ SMGs, although they represent only a modest fraction of the total high-redshift SMG population ($\simeq 10\%$). However, they have significantly higher UV luminosities, required for them to be selected as LBGs.}
\label{fig:hist}
\end{figure*}

\section{SED fitting}
\label{sec:analysis}

As noted above, \citet{Dudzeviciute_2019} have applied the {\sc magphys} energy-balance model to fit the full multi-wavelength SED of all 708 SMGs in AS2UDS.   A very useful feature of energy-balance codes such as  {\sc mapghys}, or  {\sc cigale} \citep{Noll_2009,Serra_2011,Boquien_2019}, is that the energy absorbed by dust in the rest-frame UV/optical is equated with the energy emitted in the IR. Knowing the amount of the stellar energy reprocessed by dust ($L_{\rm IR}$), one can therefore determine the level of dust attenuation in the UV through near-IR and hence, given the sufficiently deep multi-wavelength photometry, find the underlying attenuation curve and the corresponding intrinsic (before dust attenuation) stellar SED. We note that in reality the distribution of stars and dust in a given galaxy may be very complex, with multiple irregular holes in the dust and with a fraction of stars potentially situated in front of the dust. When modelling the attenuation in such a galaxy with {\sc cigale}, one assumes the shape of the attenuation curve which is in effect the mass-weighted average of the attenuation curves affecting individual regions, where the relative distribution of dust and stars is assumed to be roughly uniform.

In our analysis we adopt the photometric redshifts for the 41 LBGs in our sample from \citet{Dudzeviciute_2019}, however, as we wish to investigate the the influence of the reddening law on the relationship between the UV and IR emission of these sources, we have chosen to refit the UV and IR parts of the SEDs using a code which allows us to easily vary the attenuation curve.   Hence, we repeat the SED fitting using the energy-balance code {\sc cigale}\footnote{\url{http://cigale.lam.fr/}} \citep{Noll_2009,Serra_2011}, which enables us to explore varying the shape of the assumed attenuation curve. {\sc cigale} uses \citet{Bruzual_2003} stellar population templates, \citet{Chabrier_2003} stellar IMF and four values of stellar metallicity (0.004, 0.008, 0.02 and 0.05). A range of different star-formation histories (SFHs) was assumed, including both single- and double-burst, instantaneous, exponentially declining and continuous models (see \citealt{Koprowski_2018} for details) and the dust emission is modeled following the prescriptions of \citet{Casey_2012c}.

While we have chosen to use the photometric redshifts for the FIR-bright LBGs derived by 
\citet{Dudzeviciute_2019}, for the ALMA-faint sources, in the absence of useful  photometric constraints on their FIR emission, redshifts were determined using the {\sc eazy} template-fitting code. 
For this we used 11-band photometric coverage of the UDS ($UBVRi z JHK$[3.6][4.5]), as described in \citet{Hartley_2013} and \citet{Mortlock_2013}. All of these photometric redshifts were calibrated using $\sim$\,7,000 spectroscopic redshifts, with the resulting dispersion in $dz/(1+z)$ of $\sigma=0.018$ for {\sc eazy} (further details will be provided in Hartley et al.\ in prep.) and $\sigma=0.08$ for {\sc magphys} \citep{Dudzeviciute_2019}.

%ZZZ IRS could replace with the MAGPHYS photo-z using 13-band photometry U to 8um 
%ZZZ so the FIR-faint LBGs would have consistent photo-z?

\subsection{Attenuation curves}
\label{sec:sed}

In order to parametrise attenuation curves, {\sc cigale} adds two modifications to the standard Calzetti curve from \citet{Calzetti_2000}. The first modification is the alteration of the slope of the original Calzetti curve, normalised at $V$-band:

\begin{equation}\label{eq:att1}k_{\lambda} = k_{\lambda,{\rm Cal}}\times \left(\frac{\lambda}{550{\rm nm}}\right)^\delta,\end{equation}

\noindent where negative values of the relative slope, $\delta$, give steeper curves than that of Calzetti. The normalisation of the attenuation curve has been tied to $V$ band for historical reasons, where it has been shown that changes in $k_\lambda$ are associated with changes in optical extinction shortward of $\sim 600$\,nm, while extinction in the NIR remains relatively invariant \citep{Clayton_1988, Cardelli_1988}.

Since, by definition $k_\lambda$ is the attenuation normalised by its colour excess, $k_\lambda=A_\lambda/E(B-V)$, the modified curve from Equation\,\ref{eq:att1} has to be normalised with the modified version of the colour excess, hence:

\begin{multline}\label{eq:att2}k_{\lambda} = k_{\lambda,{\rm Cal}}\times \frac{E(B-V)_{\rm Cal}}{E(B-V)} \times \left(\frac{\lambda}{550{\rm nm}}\right)^\delta = \\
k_{\lambda,{\rm C}}\times \frac{R_V}{R_{V,{\rm Cal}}} \times \left(\frac{\lambda}{550{\rm nm}}\right)^\delta,\end{multline}

\noindent where $R_{V,{\rm Cal}}=A_V/E(B-V)_{\rm Cal}=4.05$ \citep{Calzetti_2000}. Using the above formulae one can find the dependence of the modified version of the total to selective extinction in the $V$-band, $R_V$, on the relative slope of the attenuation curve, $\delta$, to be:

\begin{equation}\label{eq:att3}R_V=\frac{R_{V,{\rm Cal}}}{(1+R_{V,{\rm Cal}})(440/550)^\delta-R_{V,{\rm Cal}}}.\end{equation}

\noindent The second modification is the addition of the UV bump:

\begin{equation}\label{eq:att4}k_\lambda = k_{\lambda,{\rm Cal}}\times \frac{R_V}{R_{V,{\rm Cal}}}\times \left(\frac{\lambda}{550{\rm nm}}\right)^\delta + D_\lambda,\end{equation}

\noindent where from \citet{Fitzpatrick_1986} the UV bump follows a Drude profile:

\begin{equation}\label{eq:att5}D_\lambda(B) = \frac{B\lambda^2\times (35{\rm nm})^2}{(\lambda^2-(217.5{\rm nm})^2)^2+\lambda^2(35{\rm nm})^2},\end{equation}

\noindent centred at 217.5\,nm (FWHM of 35\,nm), with $B$ being the amplitude of the bump.

Since we are interested in the intrinsic slopes of the attenuation curves, we do not want to normalise them by their colour excess. Instead, we decided to use the absolute attenuation in the $V$ band:

\begin{equation}\label{eq:att6}\frac{A_\lambda}{A_V}=\frac{k_\lambda}{k_V}=\frac{k_\lambda}{R_V},\end{equation}

\noindent with two free parameters: the relative slope of the attenuation curve, $\delta$, and the magnitude of the UV bump, $B$.

\subsection{Physical properties}
\label{sec:props}

We use the {\sc cigale} SED fitting procedure described above in order to derive basic physical properties for our sample: stellar masses, $M_\ast$, and UV \& IR luminosities. The best-fit values for the stellar masses are shown in Figure\,\ref{fig:hist} for the  ALMA-detected (green) and the ALMA-undetected (blue) LBG samples. In addition, we show the AS2UDS SMGs that have not been classified as LBGs (red), where the physical properties were adopted from \citet{Dudzeviciute_2019}. We allow the attenuation curve to vary, where in Section\,\ref{sec:mass} we discuss the effect it has on the resulting values of $M_\ast$. The UV luminosity at rest-frame 1600\AA\, is defined here as $L_{\rm UV}\equiv \nu_{1600}L_{1600}$, where the luminosity density at 1600\AA, $L_{1600}$, was determined from the best-fit SED (Figure\,\ref{fig:hist}).  As noted above, the photometric redshifts for the ALMA-bright sample were calculated from the rest-frame UV-radio photometry, using the high-redshift version of the {\sc magphys} code \citep*{Cunha_2008, Cunha_2015}, and are presented  in \citet{Dudzeviciute_2019}. 

To check for consistency, we compare the estimated stellar masses from our {\sc cigale} analysis to those derived using {\sc magphys} by \citet{Dudzeviciute_2019}.  The mean mass for the 41 LBGs from {\sc cigale} is ${M_\ast= 5.3\pm 1.7 \times 10^{10} {\rm M_\odot}}$, consistent with the estimate from {\sc magphys} within the 1-sigma error bars: ${M_\ast= 6.4\pm 1.4 \times 10^{10} {\rm M_\odot}}$.

\begin{figure}
\begin{center}
\includegraphics[scale=0.77]{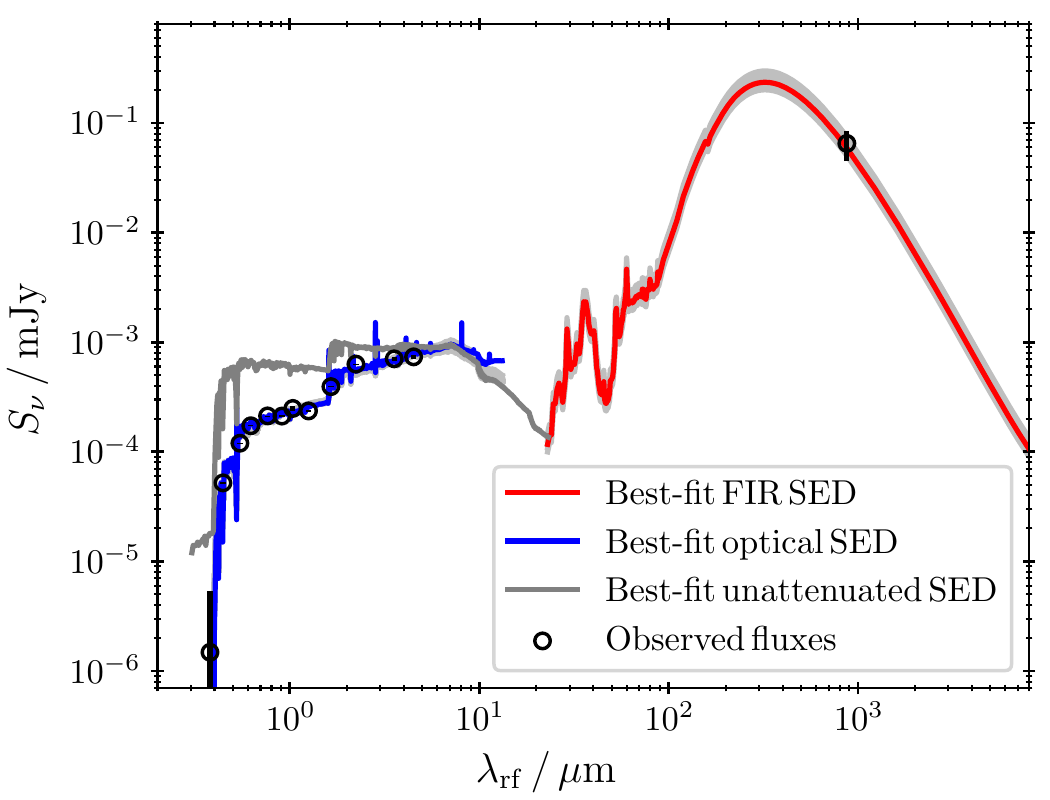}
\end{center}
\caption{Best-fitting SEDs for 209 ALMA-faint LBGs lying within the primary beam coverage of the AS2UDS ALMA maps. The rest-frame UV--NIR photometry are median values, while the FIR point represents the $3\sigma$ ALMA detection found by stacking the 209 LBGs in the ALMA maps ($S_{870}=65\pm 20\,{\rm \mu Jy}$). The resulting attenuation curve slope is $\delta=-0.10 \pm 0.25$. As explained in Section\,\ref{sec:irx2}, the relatively steep slope is consistent with the stacked population lying below the local IRX-$\beta$ relation.}
\label{fig:sedst}
\end{figure}
%ZZZ IRS how is the Td chosen for the FIR side of this fit?
%ZZZ shouldn't there be a range of similar L_IR SEDs with different Td allowed?

As explained earlier, the focus of this work is to understand the influence of the 
attenuation curve on the IRX-$\beta$ relation of high-redshift LBGs.  We are therefore 
employing the  {\sc cigale} code for the SED fitting. Producing accurate fits, however, requires a large number of SED models to be fit to the photometry, which is computationally very expensive. For this reason, we  follow \citet{Salim_2018}, who instead of fitting the multi-wavelength FIR and the UV-NIR photometry simultaneously, modified {\sc cigale}  to  include the IR luminosity as a single constraint in the energy-balance calculation.  
This approach is ideal for our analysis as we are not interested in details of the IR emission, such as characteristic dust temperature, instead our focus is on the rest-frame UV properties.
Adopting this simplified approach therefore allows us to model the stellar emission more accurately, by increasing a number of available SFHs, as well as allowing for a more detailed sampling of the dust attenuation laws.

%
%
%
% \begin{figure*}
% \begin{center}
% \includegraphics[scale=0.77]{beta_comp.pdf}
% \end{center}
% \caption{The comparison of the UV slopes, determined using four different methods, colour-coded with the UV bump strength, $B$ (Equation\,\ref{eq:att5}). The abscissa shows the values determined by fitting a power-law function to the available photometry, in the wavelength range between 125 and 200, 250 and 350\,nm (left, middle and right panels, respectively). The ordinate represents the difference between the UV slopes, as calculated by fitting a power-law to the best-fit SEDs, and the corresponding abscissa value. As can be seen from the level of scatter, as well as the size of the error bars, the accuracy of the established values increases with the breadth of the wavelength range across which $\beta$ is calculated. The   solid lines enclose the areas which are permitted, given the range of values spanned by the SED-based UV slopes (see Section\,\ref{sec:irx1} for details).}
% \label{fig:beta}
% \end{figure*}
%ZZZ IRS not convinced this plot is essential enough to include
%ZZZ can just report which range we choose and why

In order to determine simple estimates of the IR luminosity for each of the ALMA-detected LBGs, we fit the dust emission SEDs, using 185 FIR SED templates compiled by \citet{Swinbank_2014}. These include local galaxy templates from \citet{Chary_2001}, \citet{Rieke_2009}, and \citet{Draine_2007} as well as high-redshift starburst galaxies from \citet{Ivison_2010} and \citet{Carilli_2011}, with a range of dust temperatures spanning 19--60 K. We find the best-fitting SEDs using a $\chi^2$ minimization approach and determine the $L_{\rm IR}$ by integrating between the rest-frame 8--1000\,${\rm \mu m}$ emission. The resulting values are depicted in Figure\,\ref{fig:hist}, with the mean value of 
${L_{\rm IR}= 5.0^{+2.6}_{-1.5}\times 10^{12} {\rm L_\odot}}$ consistent with the {\sc magphys} derived estimates from \citet{Dudzeviciute_2019} of $L_{\rm IR}= 5.5^{+2.8}_{-1.6}\times 10^{12} {\rm L_\odot}$ (with errors representing the standard deviation), where the small difference can be attributed to the variation in the FIR libraries used.

For the ALMA-undetected sources we measure their average ALMA Band-7 flux by inverse-variance stacking in the ALMA maps. We acquire a $3.2\sigma$ detection of $S_{870}=65\pm 20\,{\rm \mu Jy}$, which is consistent with the ALMA-non-detections stack of \citet{Koprowski_2016}, but significantly lower than the $z \simeq 3$ stacks of \citet{Coppin_2015} and \citet{Koprowski_2018}. This latter discrepancy is due to two main factors. One is that in these works the LBG samples used for stacking also would have included any FIR-detected sources. The other is that our parent OIR catalogue is 0.7 mag deeper in $K$-band, which means that about a half of the ALMA-undetected sample of this work, would not have been selected using the shallower DR8 version. Since the faintest $K$-band sources are also expected to be the least massive and, therefore, the least dusty ones, our FIR stacks should be, and indeed are, fainter than those of \citet{Coppin_2015} and \citet{Koprowski_2018}. To translate the stacked flux to the IR luminosity, we adopt the $z \simeq 3$ best-fit FIR SED from \citet{Koprowski_2018}, and integrate the SED between rest-frame 8--1000\,${\rm \mu m}$ (Figure\,\ref{fig:sedst}). 

In order to determine the errors we have performed a simple Monte Carlo simulation, where we have varied the redshifts and the photometry of each source by a random offset sampled from a Gaussian distribution with $1\sigma$ equal to the corresponding errors. The $1\sigma$ systematic errors on each of the physical parameters were then found by calculating the standard deviations from modeled sources. All the resulting values are summarised in Table\,\ref{tab:prop}. The results are depicted in Figure\,\ref{fig:hist}.

\begin{figure*}
\begin{center}
\includegraphics[scale=0.77]{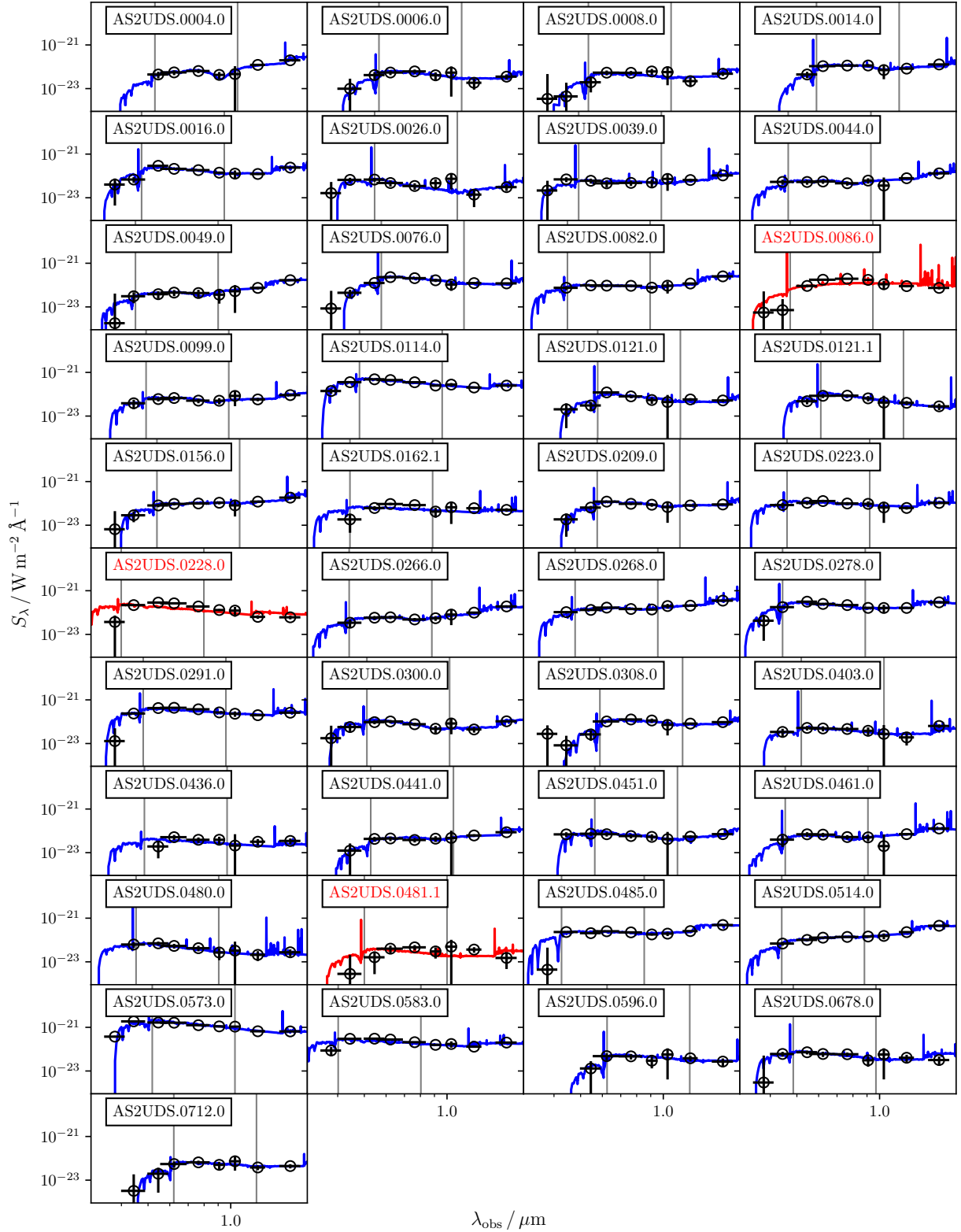}
\end{center}
\caption{Rest-frame UV part of the best-fit SEDs for the FIR-bright LBGs studied in this work (see Section\,\ref{sec:props} for details), with the UV slope range enclosed by two vertical grey lines. We mark the `bad' rest-frame UV SED fits in red.}
\label{fig:sedsall}
\end{figure*}

\subsection{Values of IRX \& \boldmath${\beta}$}
\label{sec:irx1}

To calculate IRX, we simply divided $L_{\rm IR}$ by $L_{\rm UV}$, which were measured in the previous Section. The values of UV slopes, $\beta$, can be determined in various different ways (see \citealt{Rogers_2013} for a review). In this work we decided to fit the power-law slope to the best-fitting SED between the rest-frame 125 and 250\,nm. This is because we possess only $\sim 4$ photometry points on average within the rest-frame UV range (see Figure\,\ref{fig:sedsall}) and therefore the UV slopes calculated directly from the photometry, as oppose to the best-fit SEDs, have larger uncertainties. We note that \citet{Meurer_1999} and \citet{Calzetti_2000} did not find any presence of the UV bump in their samples. We, therefore, set the amplitude of the UV bump (Equation\,\ref{eq:att5}) to 0, since it falls within the UV slope range (central rest-frame wavelength of 217.5\,nm and FWHM of 35\,nm) and could potentially affect the resulting $\beta$, by artificially shifting it towards bluer colours. The errors were calculated as in previous section.

% In Figure\,\ref{fig:beta} we plot the comparison of $\beta$'s, with the x-axis depicting the values of UV slope determined from the photometry between 125 and 200, 250 and 350\,nm (left, middle and right panels, respectively), with the y-axis showing the difference between $\beta$ found from the SED fits and determined from the photometry ($\Delta \beta=\beta_{\rm SED}-\beta_{\rm phot}$). It is clear that the narrower the wavelength range across which $\beta$ is calculated, the larger the uncertainties. While we do not know which method is most accurate, we choose to use the SED-based UV slopes, because we possess only three photometry points on average within the UV slope range, which causes the values of $\beta$ calculated from the power-law fits to the photometry to be very uncertain.

% An notable feature of Figure\,\ref{fig:beta} is the apparent anticorrelation of $\Delta \beta$ with $\beta_{125-200}$ or $\beta_{125-250}$. The reason for this is due to the way in which we chose to present the data. Since the y-axis represents the difference between the SED- and photometry-based $\beta$'s, $\Delta \beta = \beta_{\rm SED}-\beta_{\rm photo}$, the relation between $\Delta \beta$ (ordinate) and $\beta_{\rm photo}$ (x-axis) is a straight line with a slope of $-1$. Because the UV slopes calculated from SEDs range between $-1.89$ and $1.23$, all the points in Figure\,\ref{fig:beta} must lie between the two limiting lines.

It is also crucial, for the reasons explained in Section\,\ref{sec:irx2}, to determine the intrinsic (before dust attenuation) shape of the stellar emission SED, with the corresponding intrinsic UV slope, $\beta_{\rm int}$. This is important for two reasons. One is that the exact shape of the intrinsic stellar SED is required in order to translate the observed SED (given the amount of energy re-emitted by dust, $L_{\rm IR}$), into the attenuation curve. The other is that $\beta_{\rm int}$ is roughly connected to the age of the underlying stellar population, which in turn affects the position of a given source on the IRX-$\beta$ plane (see Section\,\ref{sec:irx2}). Together with the attenuation curves, the intrinsic stellar SEDs, with the corresponding values of $\beta_{\rm int}$, are determined by performing  SED fits to the available rest-frame UV-FIR photometry.

\begin{figure*}
\begin{center}
\includegraphics[scale=0.77]{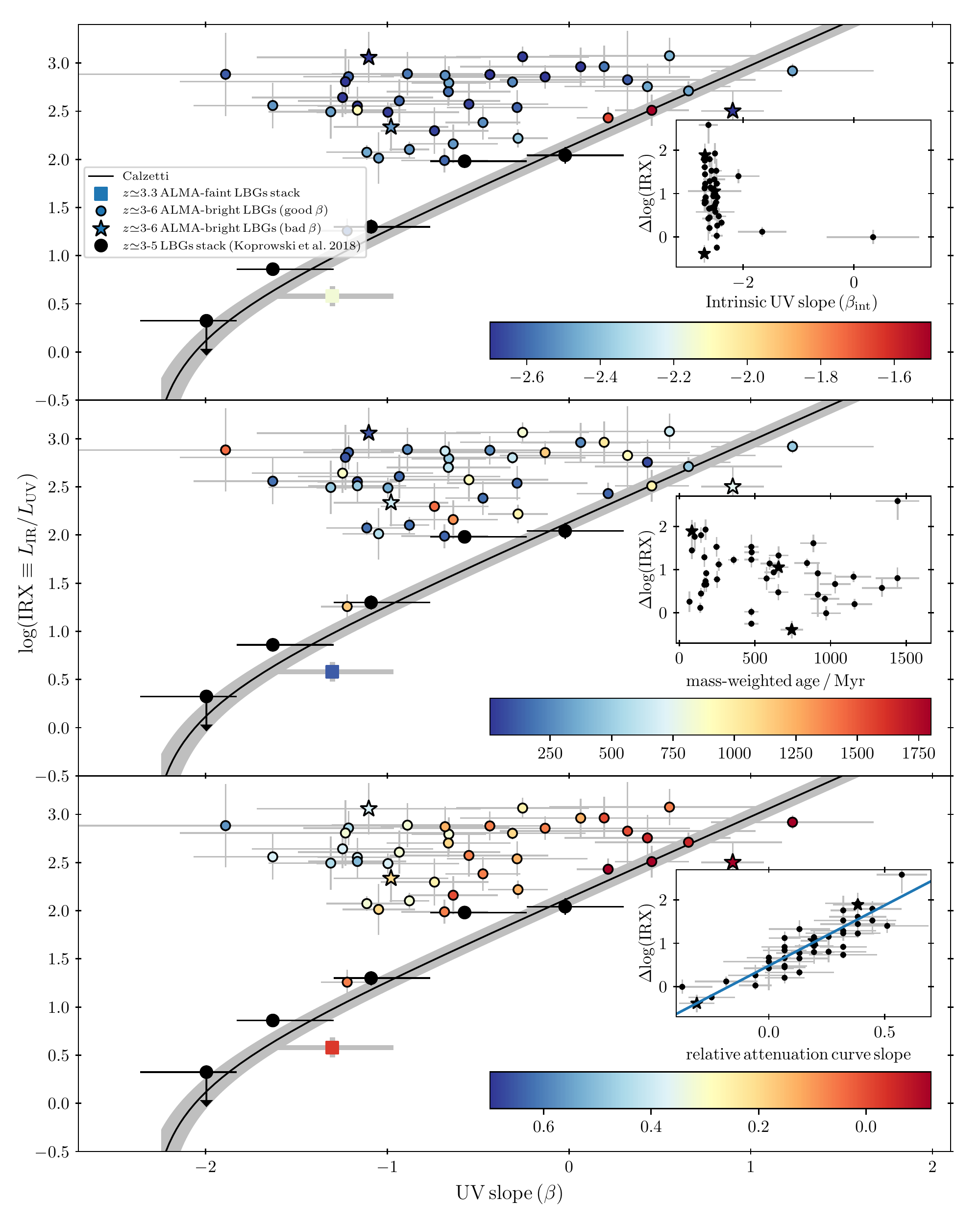}
\end{center}
\caption{Infrared excess, IRX, as a function of the UV slope, $\beta$. The FIR-bright LBGs (small circles for good $\beta$ fits and small stars for bad $\beta$ fits; see Figure\,\ref{fig:sedsall}) and the FIR-faint stack (large square) are colour-coded with the the intrinsic (before dust attenuation) UV slope, mass-weighted age and the relative attenuation curve slope, $\delta$, (top, middle and bottom panels, respectively). It can be seen in the inset plot of the bottom panel, that the distance along the ordinate of a given source from the average relation, $\Delta{\rm IRX}$, most obviously correlates with the relative attenuation curve slope, $\delta$, with the Spearman correlation of $0.89$ and the two-sided $p$-value of $\ll 0.001$, indicating significant correlation. The best-fit linear function gives $\Delta{\rm log(IRX)}=(2.79\pm 0.22)\times \delta + (0.48\pm 0.06)$. We, therefore, conclude that the scatter in the IRX-$\beta$ plane is driven mainly by the variations in the shapes of the underlying attenuation curves (see Section\,\ref{sec:irx2} for details).}
\label{fig:irxb}
\end{figure*}

%ZZZ IRS  do we think we can measure either Beta_int or mass-weighted age with
%ZZZ sufficient S/N to actually mean that the colour scales on these plots show
%ZZZ a trend??!  or are the errors so large they would wash out the trend so 
%ZZZ it wouldn't be visible.  
%ZZZ basically is the dispersion of the points / normalised by their estimated
%ZZZ errors around the mean/median >1 or <~1 ?

%ZZZ IRS the strength of the trend in the delta-beta panel is such that i think
%ZZZ this isn't real astrophysics it is simply a circular argument in the logic:
%ZZZ delta-beta is related to the offset in beta, so when you plot it against
%ZZZ that offset you get a 1:1 correlation...

%ZZZ IRS for the age/beta_int trends - is there any way to stack the SEDs of subsets
%ZZZ to demonstrate that there are/are not real trends with offset?

It is important to note, that the attenuation curve slopes and the star-formation histories (with the resulting intrinsic UV slopes) are degenerate. One can in principle increase the amount of dust attenuation by either assuming younger stellar populations (bluer $\beta_{\rm int}$), or making the attenuation curve slope shallower. However, fixing the attenuation curve to that of Calzetti will produce SED fits which have statistically higher values of $\chi^2$ (worse fits) than when the attenuation laws are allowed to vary.

Energy-balance SED fitting codes, such as {\sc cigale},  equate the amount of energy absorbed by dust in the rest-frame UV/optical bands, with the amount of energy re-emitted in the IR. Since our sources are IR-bright by selection, the SED fitting procedure will determine a need for a large amount of dust attenuation. Since, on the other hand, our LBG sample colour selection requires (by definition) detections in the rest-frame UV, it is unsurprising that the LBGs will tend to have relatively blue observed UV slopes. Applying a large amount of dust attenuation to a blue galaxy will yield a correspondingly bluer intrinsic UV slope. However, there is a limit to how intrinsically blue a galaxy can be. This limit is set by the youngest population of stars produced with the \citet{Bruzual_2003} stellar population templates and \citet{Chabrier_2003} stellar IMF. For the stellar population age of 1\,Myr, the bluest UV slope possible within the models is about $-2.7$. Fixing the \citet{Calzetti_2000} attenuation curve will, therefore, often result in the observed rest-frame UV section of the best-fit SEDs being too red, yielding poor fits to the photometry (large values of $\chi^2$), indicating that one indeed needs to allow the attenuation curves to vary in order to produce correct SED fits, or accept that the UV and FIR parts of the SED are spatially decoupled.

\section{Results \& Discussion}
\label{sec:disc}

\subsection{IRX-\boldmath${\beta}$ scatter}
\label{sec:irx2}

The main observational parameter that drives the average IRX-$\beta$ relation (Figure\,\ref{fig:irxb}) is the observed UV slope, $\beta$. To first order, the more dust attenuation in the rest-frame UV, the redder the UV slope, and the larger the corresponding value of the IR luminosity, i.e.\ IRX. However, as can be seen in Figure\,\ref{fig:irxb}, individual FIR-detected sources tend to lie above and to the left of the average relation. In order to explain the apparent scatter, we investigate the variations in two additional SED parameters, intrinsic UV slope and the dust attenuation curve shape.

\begin{figure}
\begin{center}
\includegraphics[scale=0.77]{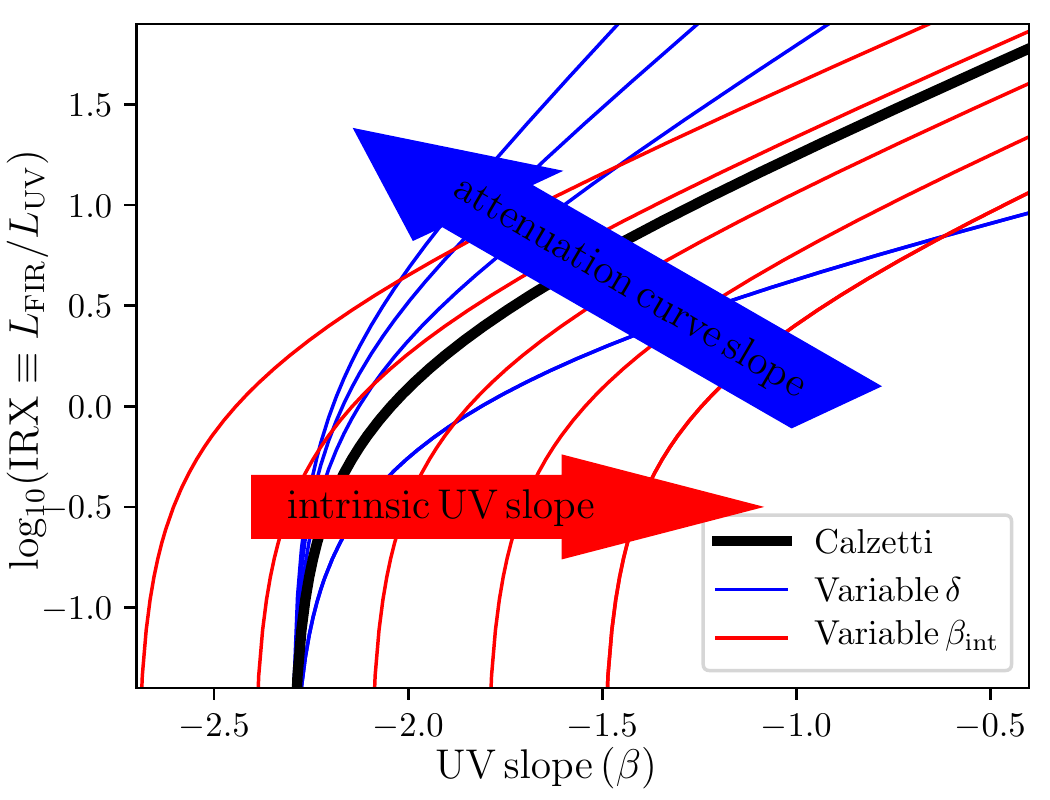}
\end{center}
\caption{The IRX-$\beta$ plane with the two main SED fitting parameters responsible for the apparent scatter shown in Figure\,\ref{fig:irxb}. While the average relation is driven by the amount of dust present in a given galaxy, the addition scatter can be introduced by either variations in the intrinsic UV slopes (red curves), or the differences in the slopes of the assumed attenuation curves (blue curve).}
\label{fig:irxbt}
\end{figure}

As can be seen in Figure\,\ref{fig:irxbt}, for a fixed attenuation curve, variations in the intrinsic UV slopes produce a scatter (red curves). This is because, at a given value of the IRX (i.e.\ given amount of the dust attenuation), different intrinsic UV slopes (value of $\beta$ for ${\rm IRX} \rightarrow 0$) will yield a variation in the measured $\beta$. Similarly, varying attenuation curve slopes will result in multiple values of $\beta$ for a given IRX (blue curves).

\subsubsection{Scatter from the intrinsic UV slopes}
\label{sec:scat1}

In the top panel of Figure\,\ref{fig:irxb} we present the scatter in the IRX-$\beta$ plane colour-coded with the rest-frame UV slope of the intrinsic (before dust attenuation) stellar emission spectrum, $\beta_{\rm int}$. It can be seen that ALMA-detected LBGs tend to have bluer $\beta_{\rm int}$ than their FIR-faint analogues. This indicates that the former sample must have a population of very young stars, responsible for the relatively blue UV slopes. However, it is known that the most massive galaxies tend to be older (e.g., \citealt{Thomas_2005}, \citealt{McDermid_2012} and \citealt{Pacifici_2013}). From Figure\,\ref{fig:hist}, it is clear that the ALMA-bright sources are much more massive than the ALMA-faint sample and so they are expected to harbour a significant population of old stars. The only way to model the SEDs with blue UV slopes and a large values of the IR luminosity is, therefore, to introduce the young burst of star formation, which is, at least in part, disconnected from the dust. The emission from these young stars will then be allowed to escape the galaxy mostly unreddened and will in effect produce relatively blue rest-frame UV SEDs. However, no clear correlation between the intrinsic UV slope and the offset between a given source and the local, average relation can be seen, which indicates that this is not the main parameter driving the scatter in the FIR-bright LBGs.

It has been proposed (e.g.\ \citealt{Kong_2004, Burgarella_2005, Boquien_2009, Grasha_2013}), that the IRX-$\beta$ scatter can be driven by the differences in the ages of the stellar populations. The stellar ages, however, affect the IRX values for a given observed $\beta$ through the effect they have on the intrinsic shape of the stellar emission SED, in particular the intrinsic UV slope, which, as shown in the top panel of Figure\,\ref{fig:irxb}, does not seem to drive the scatter above the average relation.

In the middle panel of Figure\,\ref{fig:irxb} we show the IRX and $\beta$ values for the ALMA-bright sample and the ALMA-faint stack, colour-coded with the mass-weighted age of the underlying stellar population. We assume a double-burst star formation history (SFH), where the resulting dependence of the star formation rate on time is:

\begin{equation}\label{eq:sfr} \Psi(t) \propto {\rm exp}(-t_1/\tau_1)+f_{\rm m}{\rm exp}(-t_2/\tau_2), \end{equation}

\noindent with the star forming timescale, $\tau$, for both bursts of the star formation and the mass fraction of the late burst population, $f_{\rm m}$, being kept as free parameters. This allows a large variation of the SFHs, where a number of scenarios are possible, including both single-burst and double-burst, instantaneous, exponentially declining and continuous (constant) SFHs. The effective age of a given stellar population is, therefore,  the mass-weighted mean value of the late, main burst and the young burst of star formation. As can be seen, there is no clear correlation between the mass-weighted age and the offset between a given source and the local, average relation. The likely reason for this is that the rest-frame UV range along which $\beta$ is calculated, 125 to 250\,nm, is sensitive to the most massive, young stars present in the model, which should make the intrinsic UV slope sensitive mainly to the age of the young burst of star formation, while ignoring the old burst. However, since we allow a large variations of the SFH models, the intrinsic UV slope does not even correlate with the age of the young burst alone.

\begin{figure}
\begin{center}
\includegraphics[scale=0.77]{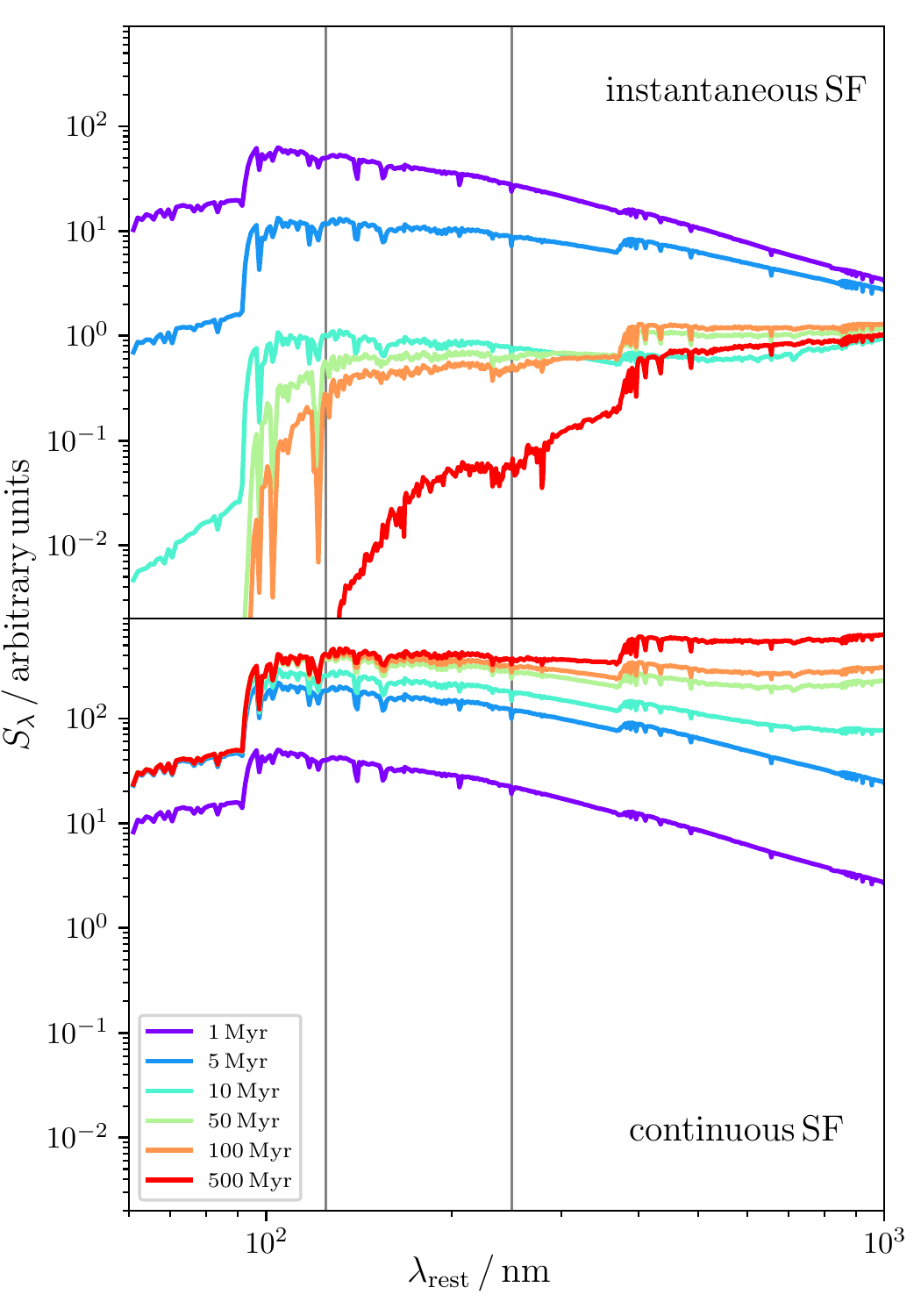}
\end{center}
\caption{The evolution of the power-law slope of the rest-frame UV part of the stellar emission spectrum with the age of the stellar population for the case of the instantaneous and the continuous star formation (top and bottom panel, respectively). The grey vertical lines represent the wavelength range along which $\beta$ is calculated (125--250\,nm). It can be seen that the form of the correlation of the intrinsic UV slope with the age of the stellar population depends on the star formation histories, which is relevant to the source of the scatter in the IRX-$\beta$ plane, as explained in Section\,\ref{sec:scat1}.}
\label{fig:seds}
\end{figure}

In Figure\,\ref{fig:seds} we show two extreme examples of the evolution of the young component of stellar SEDs with age. The top panel depicts the case of the instantaneous model, where all the stars have been formed during an instantaneous burst of star formation, while the bottom panel shows the continuous model, where stars are being formed at the constant rate. In the former model, young stars are not being replenished by the ongoing star formation, therefore the intrinsic UV slope reddens considerably quickly with age. In the latter case, the young stars are being constantly added to the population, causing the slope to redden more slowly. In addition, $\beta_{\rm int}$ is known to be sensitive to the metallicity (e.g., \citealt{Salim_2019}), with metal-poor galaxies having shallower intrinsic UV slopes. However, this effect becomes apparent only for a relatively old stellar populations and, since the mean age of the young burst in our sample, responsible for the slope of the rest-frame UV part of the stellar SED, is $25\pm 6$\,Myr, the metallicity is expected to have a negligible effect on the resulting values of the UV slope.

%ZZZ doesn't this say that beta_int (at least for a range of ages) is 
%ZZZ sensitive to SFH

\subsubsection{Scatter from the attenuation curves}
\label{sec:scatt}

\begin{figure*}
\begin{center}
\includegraphics[scale=0.77]{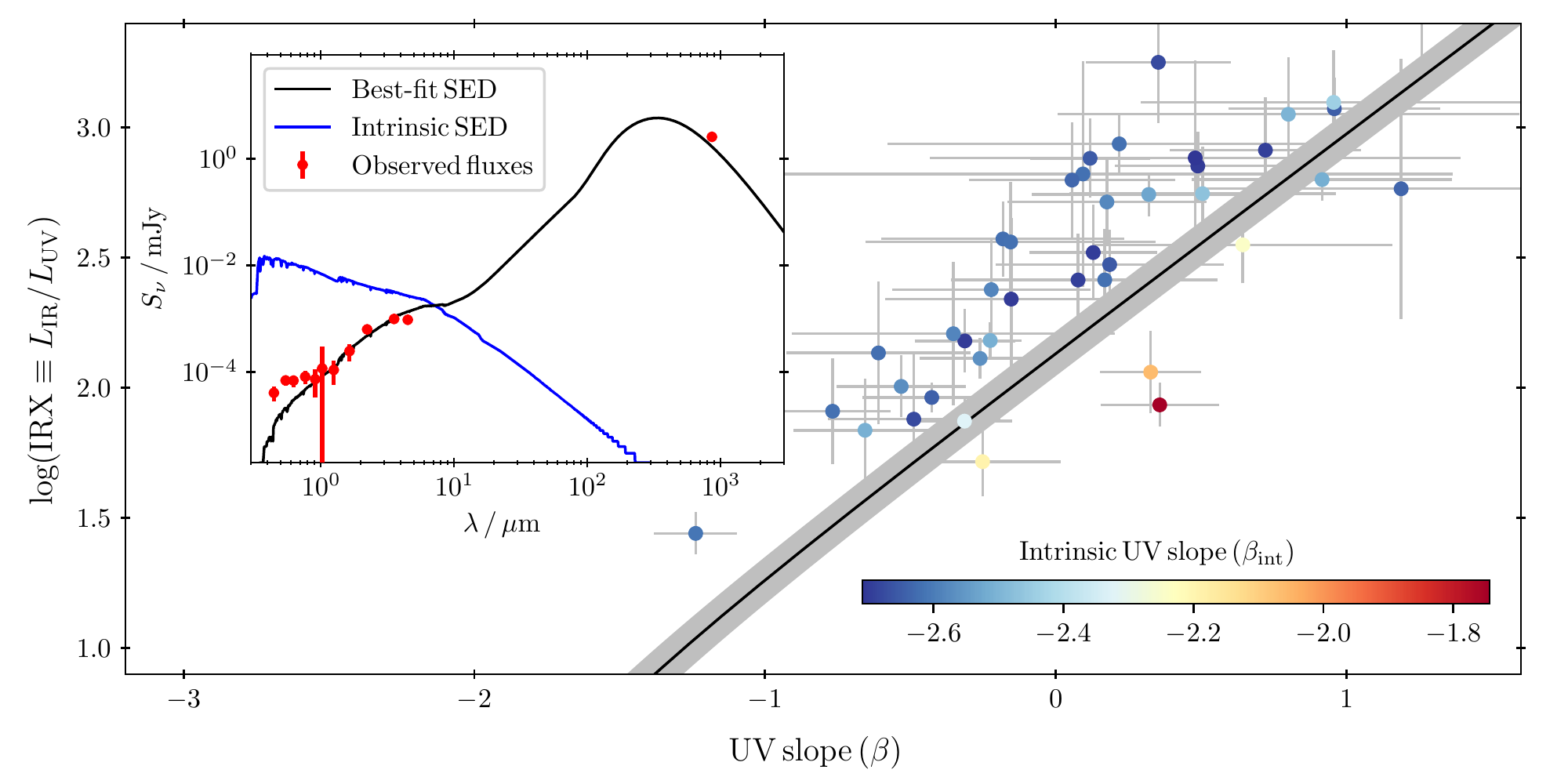}
\end{center}
\caption{IRX-$\beta$ relation, colour-coded with the intrinsic UV slope, for the ALMA-bright LBGs of this work, for the case when the attenuation law has been fixed on that of Calzetti. The inset plot shows the best-fit SED for AS2UDS.0480.0, where it can be seen that it is impossible to find a good fit without relaxing the attenuation law. The large IR luminosities of ALMA-bright LBGs forces too much dust attenuation in the optical, artificially shifting most of the sample towards redder UV slopes.}
\label{fig:irxbc}
\end{figure*}

\begin{figure}
\begin{center}
\includegraphics[scale=0.77]{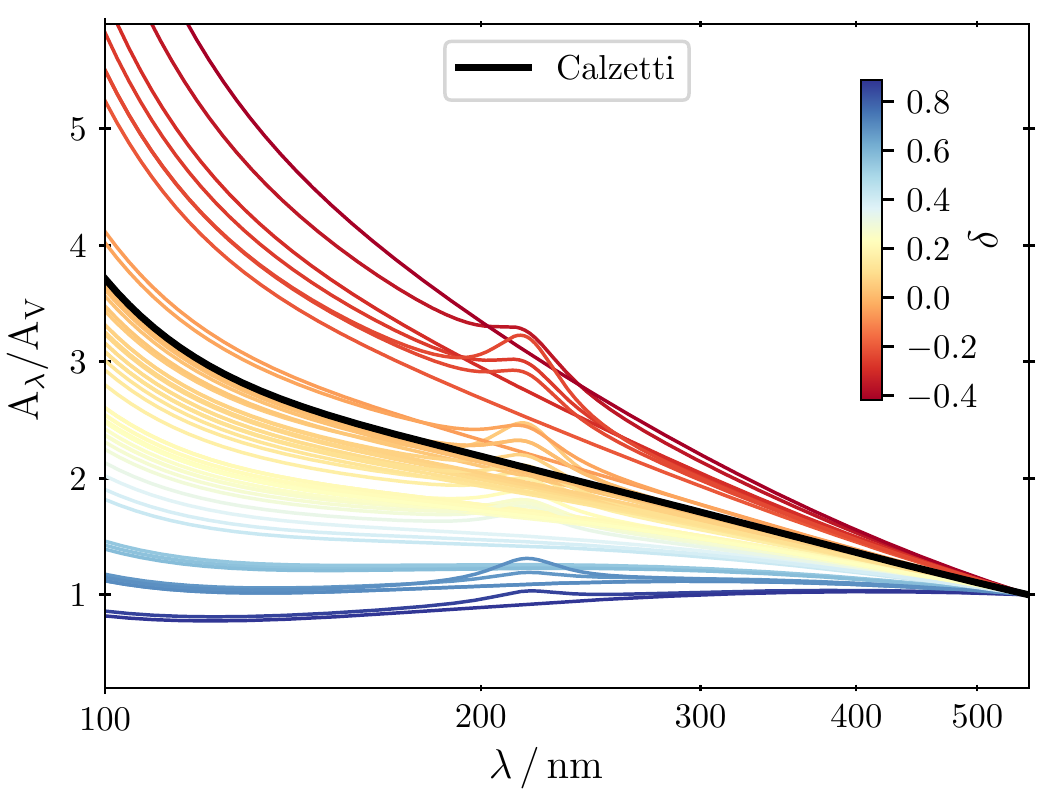}
\end{center}
\caption{Attenuation curves found for each galaxy in the sample studied in this work, colour-coded with the relative slope of the attenuation curve, $\delta$, (Section\,\ref{sec:sed}). The black thick line shows the attenuation curve of \citet{Calzetti_2000}. The range that the slopes of the attenuation curves span is about 1, while the average error is 0.1. This means that the scatter seen in the plot is real rather than a result of SED modeling uncertainties.}
\label{fig:att}
\end{figure}

A very interesting feature of the bottom panel of Figure\,\ref{fig:irxb} is that the IRX-$\beta$ scatter appears to be very well correlated with the relative slope of the attenuation curve, $\delta$ (Equation\,\ref{eq:att1}). We note here that ALMA-bright LBGs are, by selection, bright in the rest-frame UV, as well as FIR, which places them above the local IRX-$\beta$ relation. The intrinsic UV slope ($\beta$ for ${\rm IRX}\rightarrow 0$) for the galaxies lying on the local relation is $\sim$$-2.1$. Since the bluest UV slope available in our models is $\sim$$-2.7$, one can in principle, as shown in Figure\,\ref{fig:sedsall}, expect some galaxies with a very young population of stars, to end up above the local relation. However, the scatter of the intrinsic UV colors due to the age and/or metallicity for star-forming galaxies is quite small compared to the observed range, so it has a much smaller effect on the IRX-$\beta$ scatter than the attenuation curve. As shown in Fig. 6 of \citet{Salim_2019}, even for $-8.5<{\rm log(sSFR)}<-10.5$ the range of intrinsic beta is only $-2.5$ to $-2.1$ (0.4), whereas true observed slope extends over a range that is $10\times$ larger. In addition, with fixed attenuation law one simply cannot find good SED fits for most of the ALMA-bright LBGs. We show this in the inset plot of Figure\,\ref{fig:irxbc} for the AS2UDS.0480.0. The best-fit intrinsic UV slope for that source has been found to already have the minimum available value of $-2.7$. The large IR luminosity forces so much dust attenuation in the optical, that the resulting best-fit observed stellar SED is way too red, ie. {\sc cigale} simply cannot find a decent fit with the fixed attenuation law. As the consequence, the scatter in the IRX-$\beta$ plane of Figure\,\ref{fig:irxbc} (with fixed attenuation law) is artificially shifted towards redder UV slopes, as compared with the top panel of Figure\,\ref{fig:irxb}. The only way to properly model ALMA-bright LBGs is to let the attenuation law vary, in which case the sources to the left of the local relation in the bottom panel of Figure\,\ref{fig:irxb} tend to have shallower attenuation curves. The inset plot of the bottom panel in Figure\,\ref{fig:irxb} shows the relation between the vertical offset of the IRX values for FIR-bright LBGs and the average relation, $\Delta{\rm log(IRX)}$, and the relative attenuation slope, $\delta$. It can be seen that the two quantities do correlate, with the Spearman correlation of $0.89$ and the two-sided $p$-value of $\ll 0.001$. The  line in 
the inset panel in Figure\,\ref{fig:irxb} shows the best-fit linear function:

\begin{equation}\label{eq:corr}\Delta{\rm log(IRX)}=(2.79\pm 0.22)\times \delta + (0.48\pm 0.06).\end{equation}

One of the reasons behind the variations in the attenuation curves is the type of dust (namely, the distribution of the dust grain sizes) present in a given galaxy, encoded in the form of the intrinsic extinction curve. The accurate derivation of the extinction curve relies on the precise knowledge of the intrinsic stellar emission spectrum shape, as well as on the assumption that the uniform screen of dust is situated between the source of stellar emission and the observer, neither of which may be correct.  The differences in the type of dust are in effect incorporated into the resulting shape of the observed attenuation curve. Variations in the dust extinction found within the Milky Way and Small and Large Magellanic Clouds have been found to be of about two orders of magnitude at the rest-frame UV bands (e.g.\ \citealt{Gordon_2003}). Since the corresponding differences in our sample are of about six orders of magnitude (Figure\,\ref{fig:att}), it is unlikely that any possible variations in the intrinsic extinction curves will drive the massive diversity of the observed attenuation curves plotted in Figure\,\ref{fig:att} alone.

\begin{figure*}
\begin{center}
\includegraphics[scale=0.3, trim=0cm 0cm 0cm 0cm, clip]{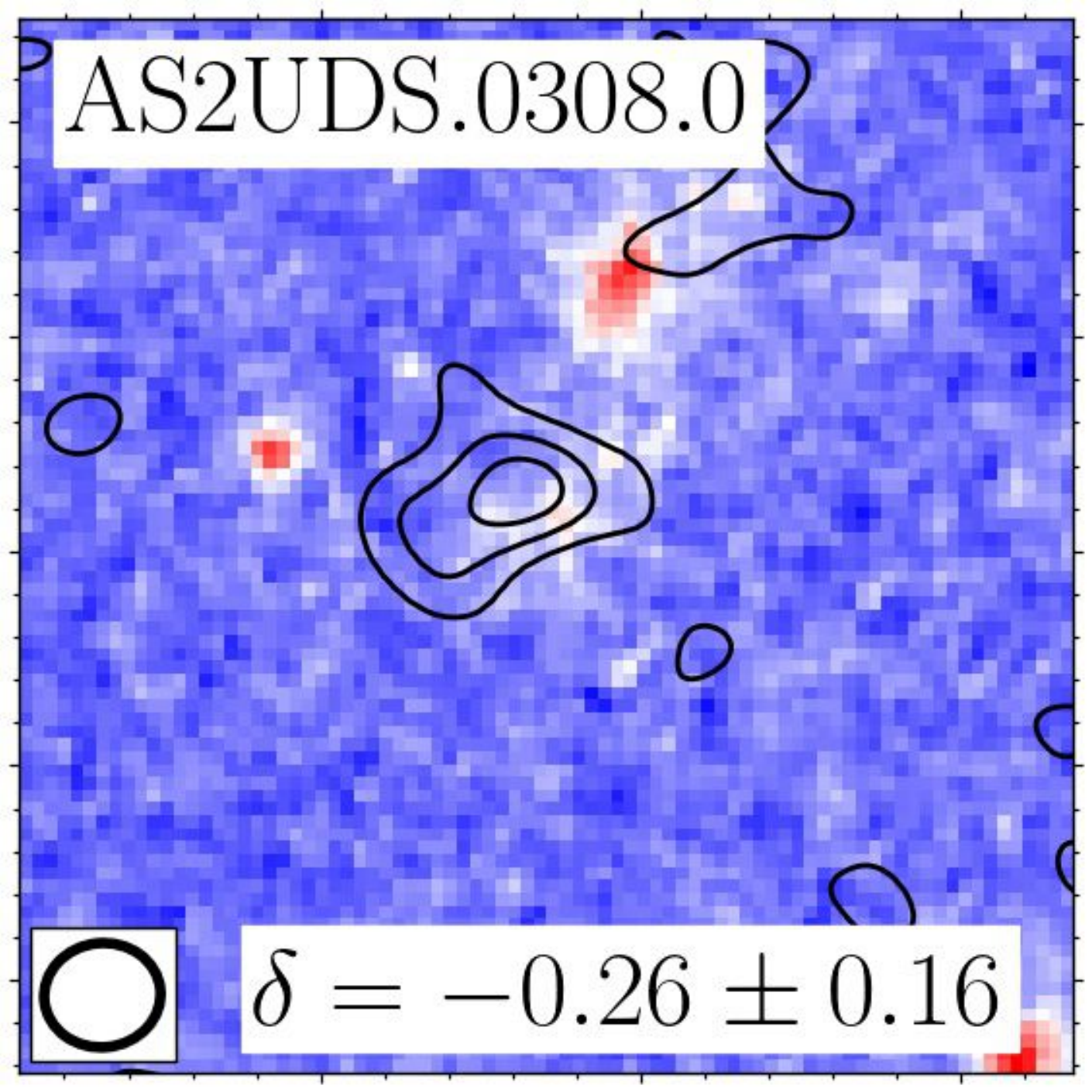}\includegraphics[scale=0.3, trim=0cm 0cm 0cm 0cm, clip]{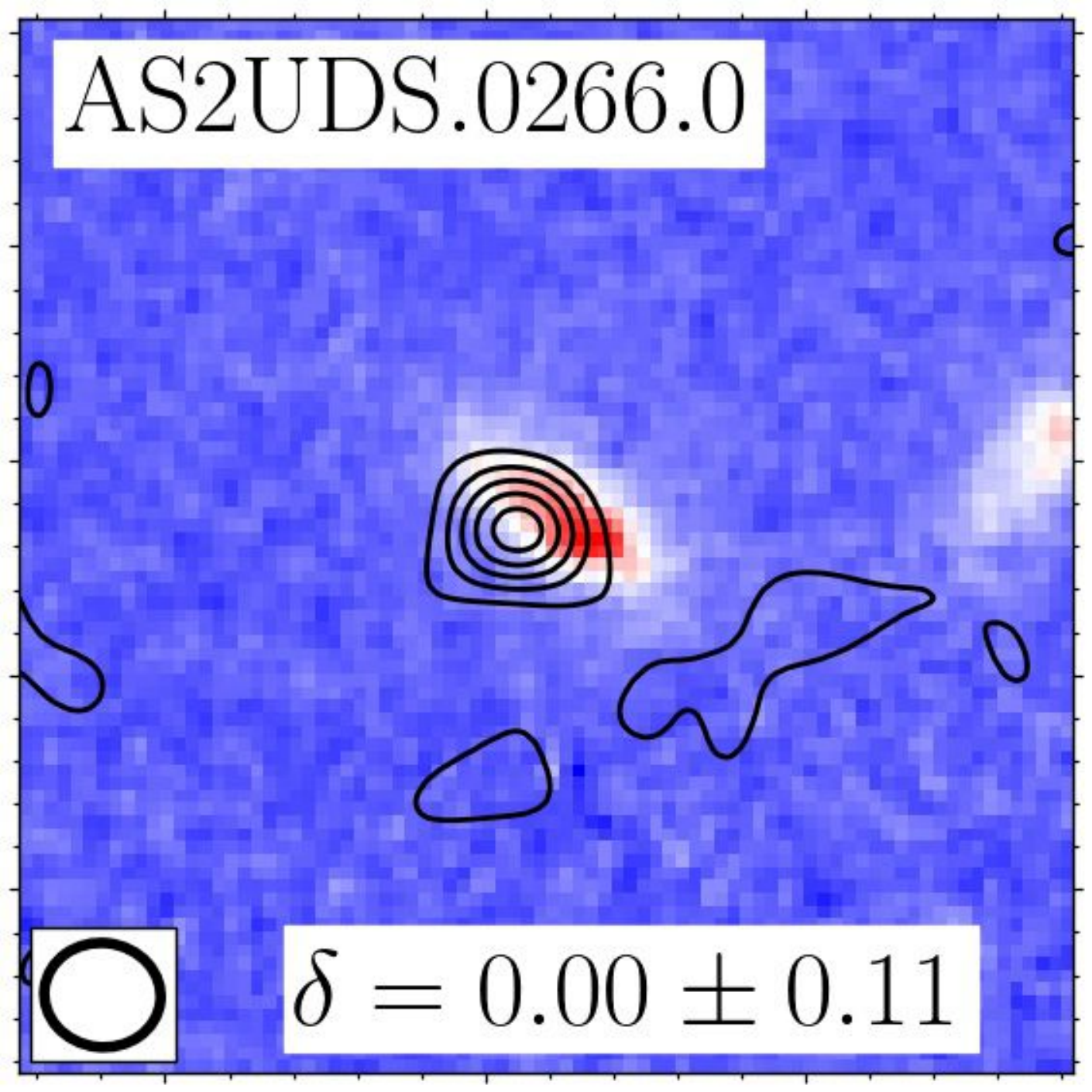}\includegraphics[scale=0.3, trim=0cm 0cm 0cm 0cm, clip]{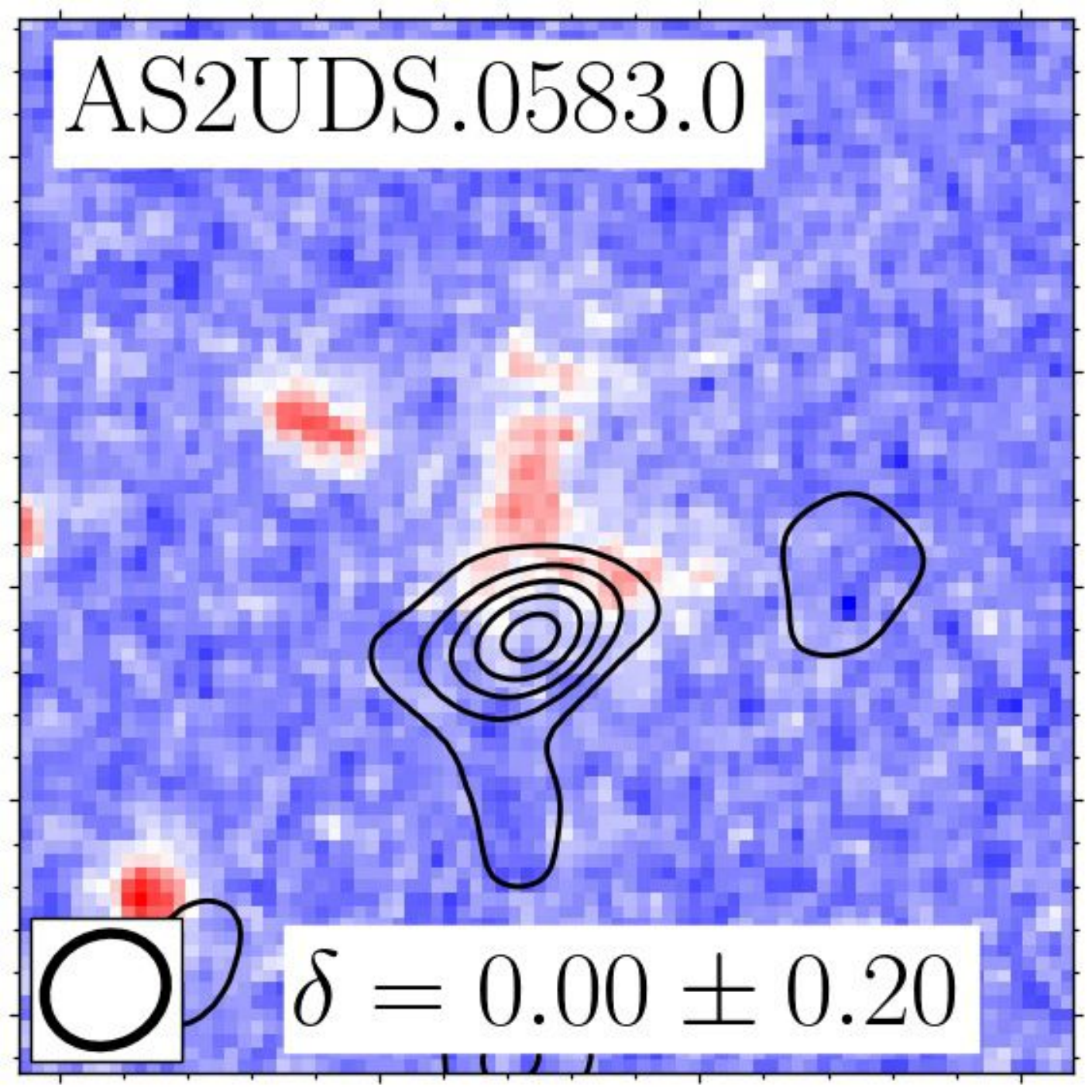}\\
\includegraphics[scale=0.3, trim=0cm 0cm 0cm 0cm, clip]{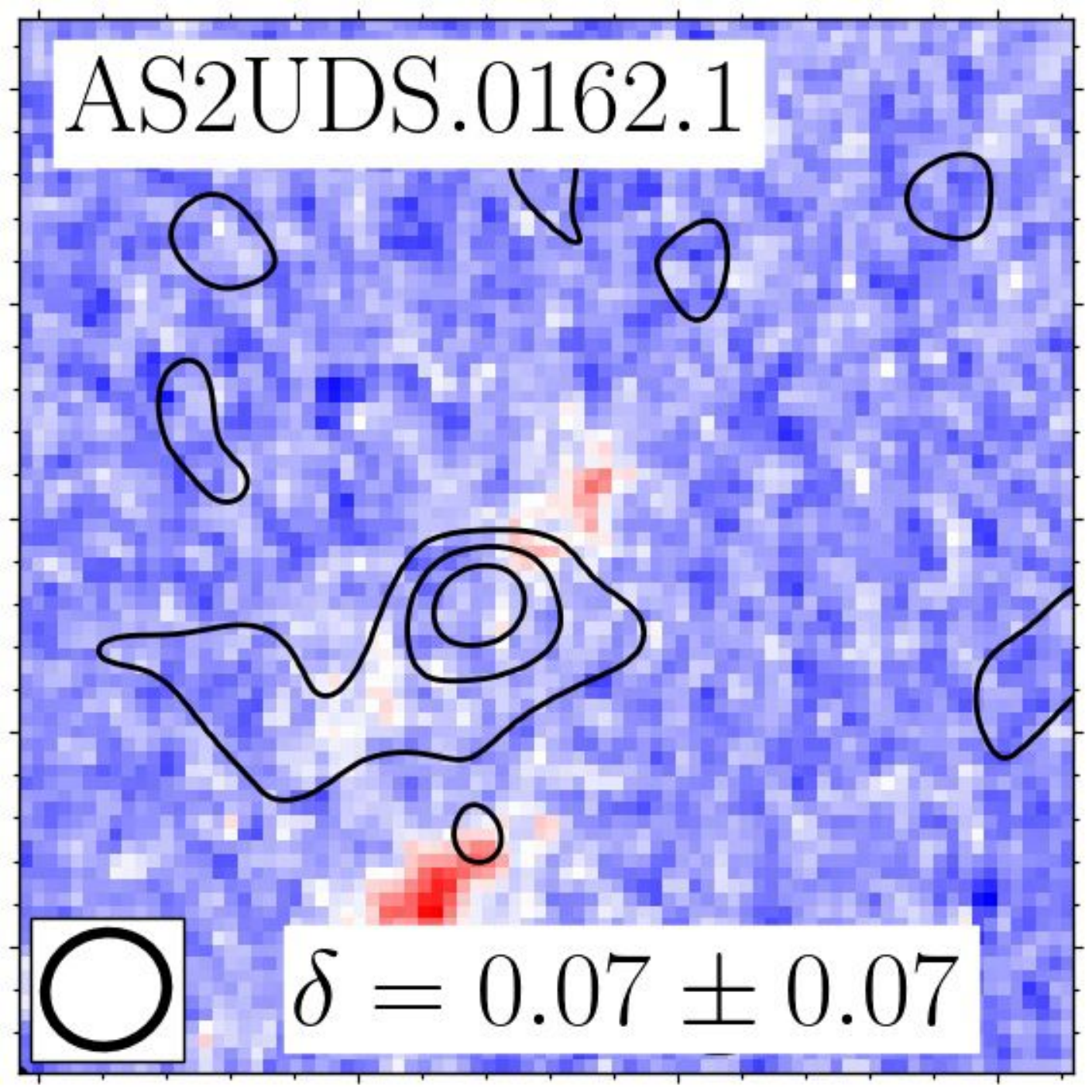}\includegraphics[scale=0.3, trim=0cm 0cm 0cm 0cm, clip]{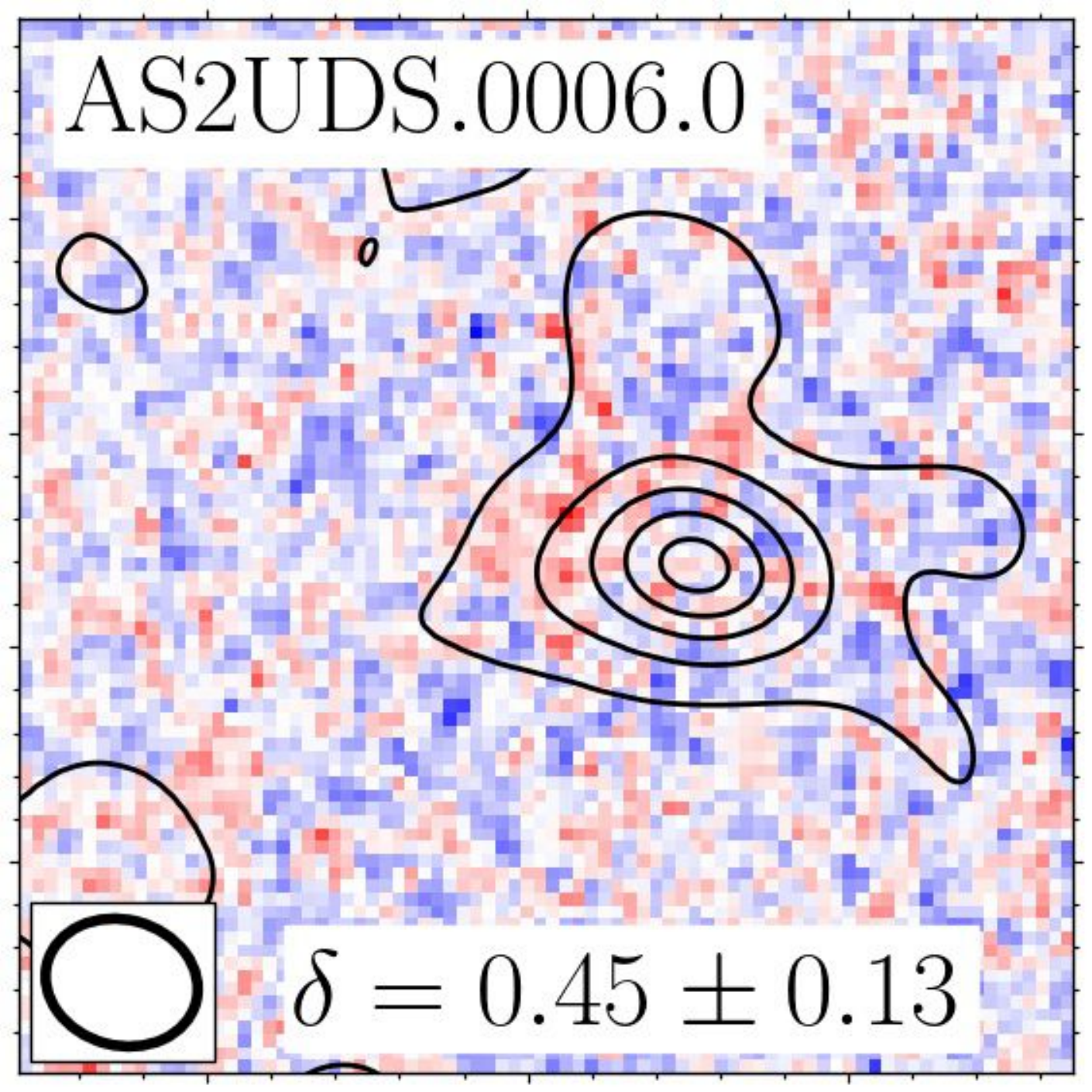}\\
\end{center}
\caption{5\,arcsec $\times$ 5\,arcsec stamps of five FIR-bright LBGs covered by the high-resolution {\it HST} imaging \citep{Stach_2019}, with the WFC3/IR F160W image in the background and the contours representing the ALMA observations ($1\sigma, 2\sigma, ...$, and in the case of AS2UDS.0006.0: $1\sigma, 5\sigma, 10\sigma, 15\sigma$ and $20\sigma$). We plot the shape of the ALMA beam in the bottom left corner and quote the corresponding values of the relative slope of the attenuation curve. The stamps are placed in the order of the increasing value of the relative attenuation curve slope. If the shape of the attenuation law depended on the level of disturbance of the stars and dust alone, one would expect it to increase as we move from left to right. Since the background image shows the stellar emission longwards of the wavelength range along which the UV slope is calculated and we only possess five such detections, no definite statements can be made on the source of the variations in the shapes of the corresponding attenuation curves. Provided that relative morphology of stars and dust is predicted to produce large range of attenuation curve slopes \citep{Ferrara_2017, Safarzadeh_2017, Popping_2017, Narayanan_2018}, we conclude that a larger sample of FIR-detected LBGs with high-resolution optical imaging would be necessary to investigate this effect more quantitatively.}
\label{fig:stamps}
\end{figure*}

As mentioned above, the age of the underlying stellar populations can drive the IRX-$\beta$ scatter through the intrinsic UV slope. It can also, however, affect it by altering the shapes of the attenuation curves (Section\,\ref{sec:irx2}). Since the star formation is triggered in the dense clouds of the ISM, the young stars ($<10$\,Myr) will be embedded in a thick shell of gas and dust (e.g.\ \citealt{Wild_2011}). One can envisage how this might manifest itself in the steepening of the galaxy attenuation curve: the younger the stellar populations, the more young stars residing in their birth clouds, and the steeper the integrated attenuation curves. This effect was investigated by \citet{Narayanan_2018b}, who found that for stellar populations younger than $\sim 2 $\,Gyr, variations in stellar ages do not significantly affect the shapes of the corresponding attenuation curves. Since our sample resides at redshifts of $z\simeq3$, the differences in the ages of stellar populations will thus likely a have minor effect on the slopes of the attenuation curves in our sample.

Another property that affects the attenuation laws in galaxies is the relative morphology of stars and dust (e.g.\ \citealt{Ferrara_2017, Safarzadeh_2017, Popping_2017, Narayanan_2018}). The original IRX-$\beta$ relationship, parametrised in the work of \citet{Meurer_1999}, was calibrated on a sample of local compact starburst galaxies, where the dust is considerably well mixed with the stars and the resulting attenuation curve was found to be consistent with that of \citet{Calzetti_2000}. Although we note, that, once corrected for far-UV aperture losses \citep{Overzier_2011, Takeuchi_2012}, the resulting Meurer relation is actually shallower than what Calzetti law predicts for IRX-$\beta$ (e.g. \citealt{Salim_2018}). However, SMGs at high redshifts have often been found to have significantly disturbed morphologies (e.g.\ \citealt{Smail_2004,Chen_2015, Chen_2017, Hodge_2018, Gomez_2018}). In this scenario, a fraction of the stellar population will be disconnected from the dust, with a small fraction of the stellar light leaving the galaxy mostly unaffected by dust; and if this exceeds the obscured emission in the UV bands then this will produce apparently very blue UV colours. On the other hand, the IRX, driven mainly by the IR luminosity, will have large values, since these sources have been selected as FIR-bright. A similar effect was observed in \citet{Wang_2018}, where the slope of the attenuation curve was affected by the galaxy inclination.

In addition, as found from the estimated dust masses and derived median hydrogen column densities, using an assumed gas to dust ratio of $\delta_{\rm gdr}=90\pm 25$ \citep{Swinbank_2014}, the SMGs in the UDS field have, on average, a $V$-band obscuration of $A_V\sim 500$ mag towards the FIR-bright component \citep{Simpson_2017}. This is in contrast with the typical $A_V$ from the SED fitting for FIR-bright LBGs in this work of order $1$--$2$ mag, which represents the dust obscuration of the stellar continuum emission detectable shortward of $\sim 1\mu$m in the restframe, integrated over the whole extent of the galaxy.
 The significant difference between these two reddening estimates suggests a real disconnect between the sites of on-going star formation (which are so highly obscured that they are optically thick into the FIR) and the more extended distribution of likely older stars which are detectable in the restframe UV and optical.  While the presence of the former component can be inferred from the FIR
 emission and so included in determining the estimated stellar sizes or mass profiles of
 the system \citep{Lang_2019}, it cannot be directly probed in the restframe UV.

 \begin{figure*}
\begin{center}
\includegraphics[scale=0.77]{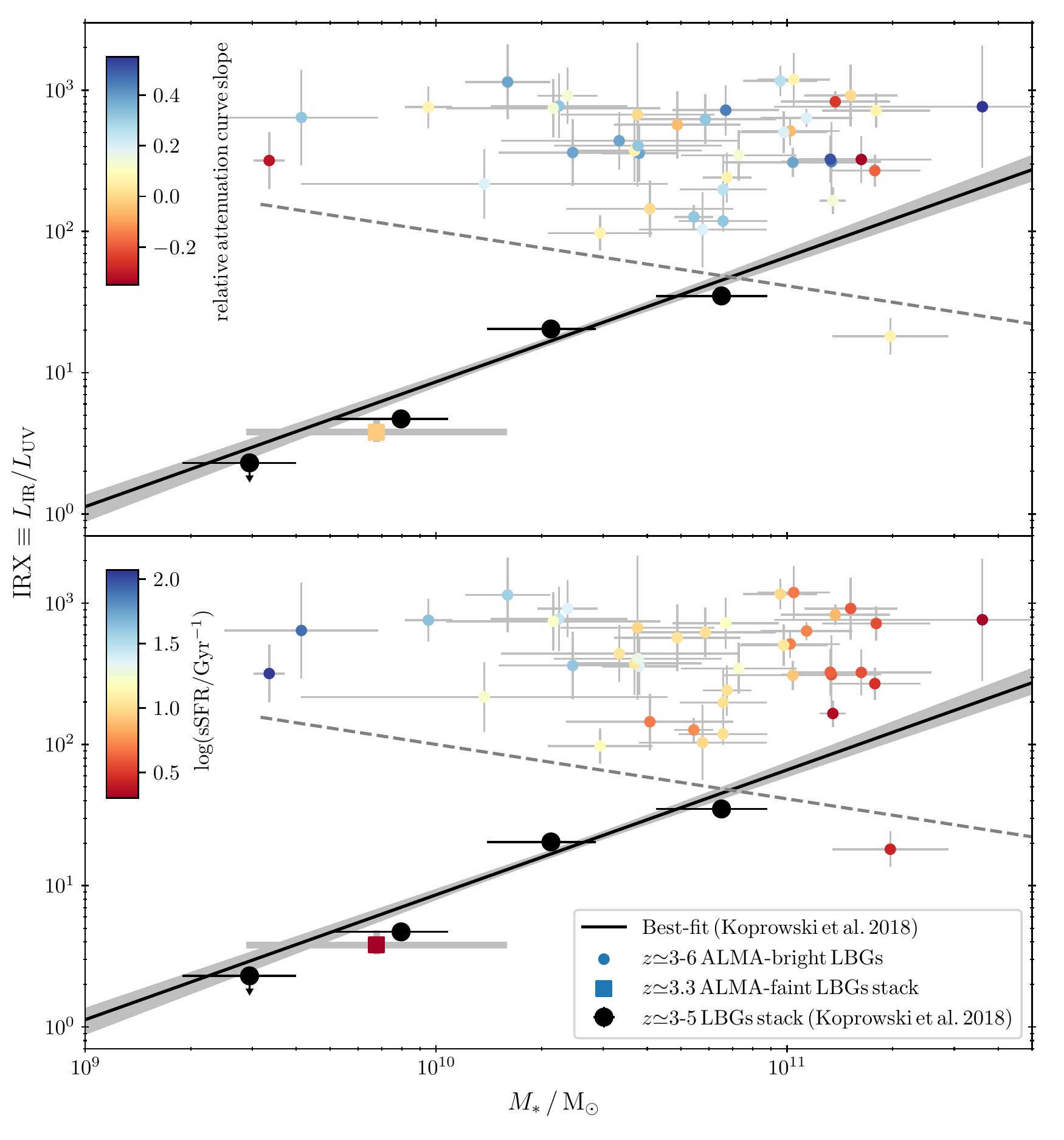}
\end{center}
\caption{IRX as a function of stellar mass, colour-coded with the slope of the attenuation curve, $\delta$ (top panel), and the specific star-formation rate, sSFR (bottom panel). The filled black circles are the stacking results for the combined high-redshift ($z=3$--$5$) LBG samples of \citet{Koprowski_2018}, fitted with a best-fit power-law. The large square shows the FIR-faint ALMA sample of this work. The grey dashed line depicts the rough selection limit for the AS2UDS sample, corresponding to the typical UV luminosity of $\sim 1\times10^{10}\,{\rm L_\odot}$ at the stellar mass of $\sim 1\times 10^{10}\,{\rm M_\odot}$, rising to $\sim 3\times10^{10}\,{\rm L_\odot}$ at $\sim 2\times10^{11}\,{\rm M_\odot}$ and the sensitivity limit of the AS2UDS of $1\times10^{12}\,{\rm L_\odot}$. It can be seen that, as opposed to the IRX-$\beta$ plane, the scatter is driven by the variations in the values of the sSFRs, with FIR-bright LBGs lying in the `starburst' regime, indicating that they could be undergoing a major merger or secular processes responsible for an enhanced phase of star-formation activity. This supports the scenario in which ALMA-detected LBGs have their shallow attenuation curves produced by a disturbed stellar morphology, likely the result of an enhanced period of vigorous star formation activity.}
\label{fig:irxm}
\end{figure*}

Hence it is expected, provided that the relative morphology of dust and stars is the driving factor of the IRX-$\beta$ scatter, that sources with flat attenuation curves (blue curves in Figure\,\ref{fig:att}), lying considerably above the local IRX-$\beta$ relation (blue points in the bottom panel of Figure\,\ref{fig:irxb}), would be expected to have, on average, relatively disturbed morphologies, while the galaxies below the local relation would be expected to have less disturbed disk-like structures. To investigate this we require high-resolution optical {\it HST} data, which we only possess for the small fraction of the UDS field covered by the {\it HST} Cosmic Assembly Near-infrared Deep Extragalactic Legacy Survey (CANDELS; \citealt{Grogin_2011, Koekemoer_2011}).

In Figure\,\ref{fig:stamps} we show the optical stamps from \citet{Stach_2019} for five IR-bright LBGs covered by the CANDELS data, where we show them in the order of increasing value of the relative slope of the attenuation curve, $\delta$, (decreasing slope of the attenuation curve, see Figure\,\ref{fig:att}). Ideally, we would like to use resolved imaging from the UV slope range of 125 to 250\,nm, in order to construct the resolved map of $\beta$, but since our sample is very dusty, they are undetected in the rest-frame UV {\it HST} imaging. Based on this small sample we can see no clear correlation between $\delta$ and the level of disturbance. However, other works seem to confirm that SMGs do tend to have disturbed optical morphologies (e.g.\ \citealt{Chen_2015, Chen_2017, Hodge_2018, Lang_2019}). Since {\it HST} WFC3/IR F160W imaging, at the redshift of our sample of $z\simeq 3$, samples the restframe near-UV it is sensitive to a mix of moderately obscured young and intermediate age stars, and we only possess five such detections, no significant statements can be made about the impact the relative morphology of stars and dust have on the shape of the resulting dust attenuation curves for FIR-bright LBGs. We conclude that a much larger sample of FIR-bright LBGs, with the high-resolution rest-frame UV to FIR data, is required in order to investigate this issue properly.

%ZZZ IRS i am sure there is restframe UV coverage of these galaxies - not just F160W
%ZZZ if you look at Stach19 thumbs then i think what you would conclude is that
%ZZZ any UV light you are seeing doesn't really reflect what is going on in these
%ZZZ galaxies and in some cases you ask what exactly we are managing to measure with
%ZZZ beta as there doesn't seem to be any blue light at all?!?!?

\subsection{IRX-\boldmath${M_\ast}$ scatter}
\label{sec:irxm}

It is often argued that the stellar mass should be used, instead of the UV slope, in order to calibrate a relationship for the IRX. This is motivated by the claims that the total stellar mass  provides an indirect tracer of the amount of the dust attenuation (e.g.\ \citealt{Sobral_2012, Whitaker_2012, Heinis_2013, Alvarez_2016, Dunlop_2017, Reddy_2018, McLure_2018}). This is because, to first order, the more massive the galaxy means that more stars will have formed (higher SFR) and more dust will be present in the ISM (larger $L_{\rm IR}$), producing redder UV colours. In Figure\,\ref{fig:irxm} we plot the values of the IRX and stellar mass for high-redshift LBGs, colour-coded with the relative slope of the attenuation curve, $\delta$, and the specific star-formation rate, sSFR,  with the average relation from \citet{Koprowski_2018} overplotted. It is clear that the FIR-bright LBGs, as in the case of the IRX-$\beta$, do not follow the average relation.

The total SFR is a sum of the dominant IR obscured SF and (negligible) observed unobscured UV SF contributions. In order to find the values for the SFRs, we follow \citet{Madau_2014}. For UV SFRs we multiply $L_{\rm UV}$ by $2.5\times 10^{-10} \,{\rm M_\odot\,yr^{-1}\,L_\odot^{-1}}$, while for IR SFRs, we multiply $L_{\rm IR}$ by $1.73\times 10^{-10} \,{\rm M_\odot\,yr^{-1}\,L_\odot^{-1}}$. Since these conversions assume \citet{Salpeter_1955} IMF, we also multiply both SFRs by a factor of 0.63 to convert to \citet{Chabrier_2003} IMF.

A clear scatter can be seen in Figure\,\ref{fig:irxm}, with the FIR-bright LBGs exhibiting significantly larger values of the IRX, given their stellar masses, than their FIR-faint analogues. Moreover, the scatter in the IRX-$M_\ast$ plane does not seem to be driven by the shape of the attenuation curves, as in the case of the IRX-$\beta$, but by the variations in the sSFRs. This is, of course, to be expected, since IRX-$M_\ast$ relationship is analogous to that between the SFR and stellar mass. 

The relation between the stellar mass and the total SFR is quantified via the so-called star-formation `main sequence' (MS; e.g.\ \citealt{Noeske_2007, Michalowski_2012a, Speagle_2014, Koprowski_2016b, Michalowski_2017}). At a given redshift, normal star-forming galaxies form a broad trend between star-formation rate and stellar mass, whereas `starbursts' are offset towards higher sSFRs by a factor of $>$\,2--4 (e.g.\ \citealt{Daddi_2007, Elbaz_2011, Speagle_2014}, although see \citealt{Elbaz_2018}). For the FIR-bright LBGs, the relatively large values of the IRX, given their stellar masses, are well correlated with the sSFRs, which indicates that these sources are undergoing a burst of SF. Again, since we possess high-resolution optical imaging only for 5 of our sources, we conclude that a larger sample with high-res UV-FIR imaging is required in order to reach any quantitative conclusions.

\subsection{Fraction of obscured star formation}
\label{sec:fobs}

\begin{figure}
\begin{center}
\includegraphics[scale=0.77]{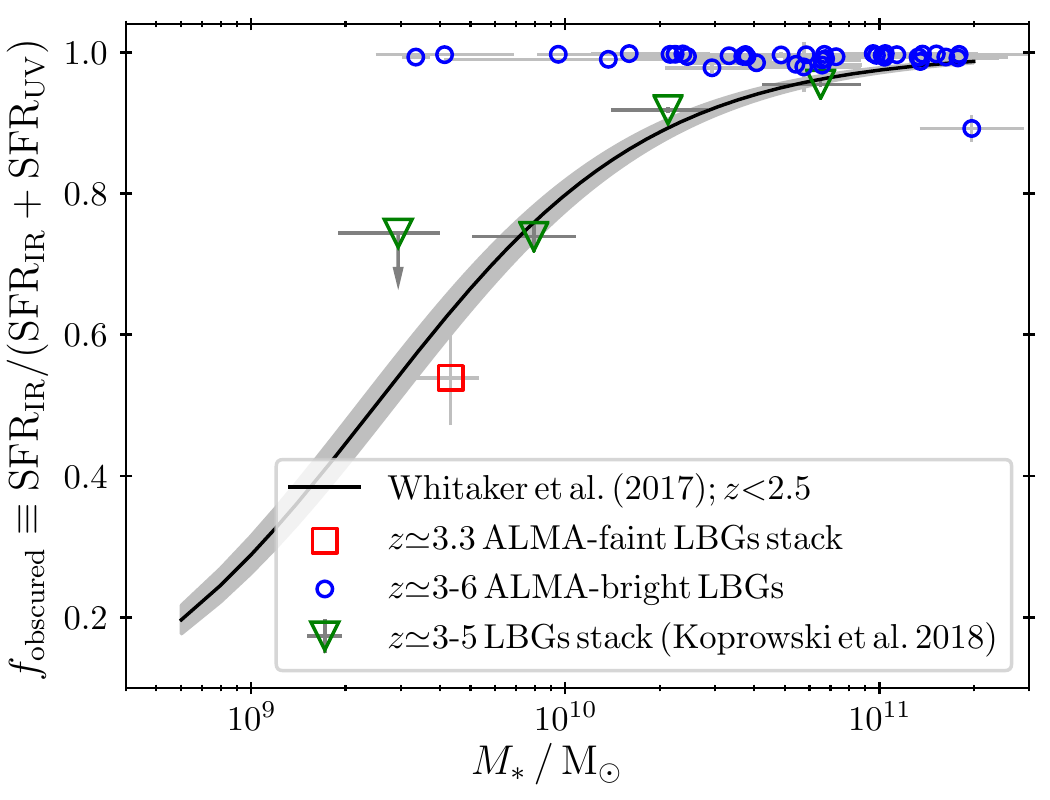}
\end{center}
\caption{The fraction of obscured star formation, $f_{\rm obs}\equiv {\rm SFR_{IR}/SFR_{UV+IR}}$, as a function of stellar mass. The solid line is the best-fit function found by \citet{Whitaker_2017}. The green open triangles in the bottom panel represent the stacking results for the combined $z\sim3$, $z\sim4$ and $z\sim5$ LBG samples from \citet{Koprowski_2018}, the red open square shows the FIR-faint ALMA sample of this work, while the open blue circles depict the FIR-bright LBGs. It can be seen that the SFRs for the FIR-bright LBGs are almost exclusively dominated by the dust component.}
\label{fig:fobs}
\end{figure}

We start by noting that  the quartile redshift range spanned by the 41 ALMA-detected LBGs analysed here corresponds to $z\sim 2.7$--3.5, but that  these LBGs comprise only $16\pm 2$\% (41 of 250) of the ALMA-detected  population in AS2UDS in that redshift range from \citet{Dudzeviciute_2019}.   Hence, this UV-bright subset is far from representative of the bulk of highly star-forming galaxies at these redshifts.  Indeed, \citet{Dudzeviciute_2019} find that
$\sim 17$\% of the SMG population are undetected down to the  $K$-band limit of $K=25.7$  of this study, most of which are expected to lie at $z\sim2$--4, demonstrating the incompleteness of even NIR-selected samples against the most active massive, dust-obscured galaxies at $z\gg2$.

As found by \citet{Whitaker_2017}, the fraction of obscured star formation, $f_{\rm obs}\equiv {\rm SFR_{IR}/SFR_{UV+IR}}$, is a strong function of the stellar mass, at least out to $z=2.5$, where most massive sources ($M_\ast> 10^{11}\,{\rm M_\odot}$) have their SFRs dominated by the dust component ($f_{\rm obs}\sim 1$). They also found that the relation does not undergo any redshift evolution. Since our sample lies at $z\simeq 3$, we decided to investigate how it compares with the lower-redshift sources of \citet{Whitaker_2017}.

We present our results in Figure\,\ref{fig:fobs}. The solid line shows the best-fit function of \citet{Whitaker_2017}. We show the stacking results for the $z\simeq 3$--$5$ LBGs sample of \citet{Koprowski_2018} and the stack of the ALMA-faint LBGs lying within the ALMA maps investigated in this work, as well as the ALMA-detected LBG sample. It is clear that both stacked LBG samples are consistent with the findings of \citet{Whitaker_2017}. This confirms that, on average, UV/optically-selected star-forming dropout galaxies have a strong relation between the fraction of obscured star formation and the stellar mass, which continues out to $z\sim3$ and beyond. However, the ALMA-bright LBGs do not follow this trend. As in the case of IRX-$M_\ast$ relation, the FIR-bright LBGs, having their SFRs dominated by the IR-luminous dust component ($f_{\rm obs}\gtrsim 0.95$), lie well above the average curve (which the more typical FIR-faint LBGs tend to follow).

\subsection{Stellar masses}
\label{sec:mass}

\begin{figure}
\begin{center}
\includegraphics[scale=0.77]{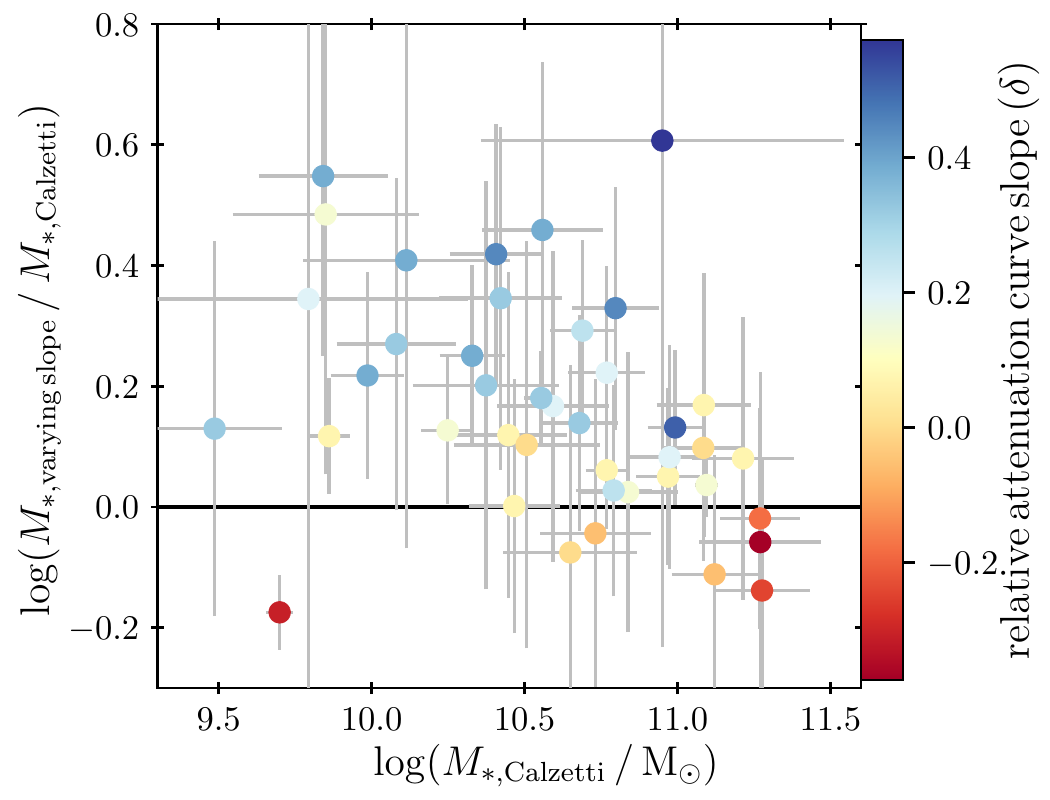}
\end{center}
\caption{Ratio of the stellar masses derived when allowing the attenuation curves to vary freely, to values determined assuming the Calzetti law (ordinate), as a function of the Calzetti-like stellar masses (abscissa), colour-coded with the relative slope of the attenuation curve, $\delta$, (see Figure\,\ref{fig:att}). It can be seen that, since fixing the attenuation curve on that of Calzetti tends to produce worse fits (larger values of $\chi^2$), assuming the Calzetti attenuation curve will often underestimate stellar masses for FIR-bright sources (see Section\,\ref{sec:mass} for details).}
\label{fig:mass}
\end{figure}

A very important consequence of the variation in the shapes of the attenuation curves is the fact that it affects the resulting values of the stellar mass. The procedure of determining stellar masses from the SED fitting relies on the translation of the observed best-fit SED into the intrinsic, dust-unaffected spectrum. The wavelengths which are driving stellar masses are the NIR wavelengths, since it is the old, evolved stars, that dominate the mass budget of a galaxy. The \citet{Calzetti_2000} attenuation curve is characteristic of the average population of LBGs (although see \citealt{Boquien_2012}, \citealt{Schaerer_2013} and \citealt{Reddy_2018}), while, as shown in Figure\,\ref{fig:irxb}, sources well above the local IRX-$\beta$ relation will have greyer (less steep) attenuation curves. For a fixed amount of energy re-emitted by dust in the IR, $L_{\rm IR}$, the amount of the stellar light attenuated by dust will be dominated by the most energetic rest-frame UV wavelengths. Altering the slope of the attenuation curve will have little effect on the dust obscuration at the rest-frame UV, since $L_{\rm IR}$ must be conserved. However, greyer curves will have lower ratio of UV-to-NIR relative attenuation and, since the dust obscuration at UV is rather insensitive to the slope of the attenuation law, shallower curves will produce higher intrinsic NIR fluxes, increasing the resulting stellar masses.

Physically, for a given observed UV--NIR photometry and fixed $L_{\rm IR}$, the age of the stellar population will increase with the shallowing of the attenuation curve, since shallower curves produce redder intrinsic UV slopes (e.g. \citealt{Leja_2019}). In this scenario more stars (i.e.\ higher stellar mass) will be required in order to produce a given IR luminosity. We show this in Figure\,\ref{fig:mass}, where the difference between the stellar masses calculated allowing the attenuation curves to vary and fixing them at the \citet{Calzetti_2000} shape is plotted as a function of the Calzetti-like stellar mass, colour-coded with the relative slope, $\delta$. Bearing in mind that fixing the attenuation curve on that of Calzetti tends to produce worse SED fits (larger values of $\chi^2$), it is clear that galaxies with the greyest curves (largest $\delta$) will tend to have their stellar masses underestimated by a factor 2$-$3$\times$, when Calzetti curve is incorrectly assumed. Since, as found by \citet{Casey_2014b}, sources well above the local IRX-$\beta$ relation, suffering from the greyest attenuation curves, have the largest values of the IR luminosity, this effect will be most prominent in the case of FIR-bright LBGs and SMGs.

%ZZZ IRS this is not right - you don't need MORE stellar mass to increase L_FIR you just
%ZZZ need to add a few more bright/young stars!

%
%
%
\section{Summary}
\label{sec:summ}

We exploit the large AS2UDS ALMA survey of the UDS field to  identify and study a  sample of 41 rare sub-millimetre-detected LBGs. We have performed a detailed analysis, including SED fitting to the rest-frame UV--FIR photometry, in order to  investigate their relation to the typical LBG samples at high-redshifts in terms of the IRX, UV slope and stellar mass. We summarise the main conclusions from our work as follows:

\begin{enumerate}

\item{We find that the FIR-bright LBGs are on average significantly more massive than their FIR-faint analogues, which supports the claimed correlation between the stellar mass and the SFR (e.g.\ \citealt{Speagle_2014}).}
%ZZZ IRS BUT NOT COMPLETE IN SFR OR MASS SO DON'T THINK WE CAN CONCLUDE ANYTHING!

\item{When performing the rest-frame UV-FIR SED fits to the available broad-band photometry, we found that the attenuation curves must be allowed to vary in order to produce reliable fits. With the individual values of the IRX and $\beta$ found from the best-fit SEDs for each source in our sample, we found that FIR-bright LBGs lie well above the average Calzetti-like IRX-$\beta$ relation, characteristic of more typical high-redshift LBGs. We confirm that, while the differences in the ages of the underlying stellar populations play a role, the scatter is mainly driven by the variations in the shapes of the attenuation curves, where the  offset between a given galaxy and the average relation correlates tightly with the attenuation curve slope, with Spearman correlation of $0.89$ and the two-sided $p$-value of $\ll 0.001$. We find the relation to be of a linear form with $\Delta{\rm log(IRX)}=(2.79\pm 0.22)\times \delta + (0.48\pm 0.06)$.}
%ZZZ IRS is this result more than just trivial?!?!

\item{We find the attenuation curves for FIR-bright LBGs to be varying by over six orders of magnitude in the rest-frame UV (see also \citealt{Dudzeviciute_2019}). While this in principle should mainly be driven by the level of relative disturbance of the stars and dust in a given galaxy (e.g. \citealt{Chen_2015,Chen_2017}), the differences in the dust type, manifested in the shapes of the intrinsic extinction curves, as well as different ages of the stellar populations (fraction of young stars residing in their birth clouds), cannot be ignored. We investigate how the shape of the attenuation curve may relate to the relative morphology of stars and dust using high-resolution {\it HST} data. However, with only five sources detected, we cannot find any evidence for a strong correlation and suggest that larger sample may be required.}

\item{We investigated the relationship between the IRX and $M_\ast$ for our sample and found that FIR-bright LBGs exhibit unusually large values of IRX, given their stellar mass, relative to the FIR-faint sources. In addition, we show that the scatter in the IRX-$M_\ast$ plane is driven by the variations in the sSFRs, where FIR-bright galaxies tend to exhibit large values, placing them in the `starburst' regime of the star-formation main sequence. This indicates that the LBGs detected in the FIR with ALMA are massive galaxies, with apparently enhanced star-formation rates, potentially driven by mergers (see also \citealt{Dudzeviciute_2019}). This supports a scenario in which the unusually shallow slopes of the attenuation curves, detected in these sources, are largely produced by the disturbed morphologies.}

\item{When stellar masses are to be determined from the SED fits, caution must be taken when assuming the shapes of the underlying attenuation curves. As found in this work, since the IR-brightest galaxies lie well above the local IRX-$\beta$ relation, then attenuation curves much greyer than that of Calzetti must be assumed, in order not to underestimate their stellar masses.}

\end{enumerate}

The FIR-bright LBGs are a SFR-incomplete sub-sample of LBGs, which by selection have both blue UV slopes and high IR luminosities, placing them above the local IRX-$\beta$ relation. We found that assuming relatively flatter attenuation curves is required in order to produce good SED fits, where the slope of the attenuation law correlates with the scatter in the IRX-$\beta$ plane. It is, therefore, incorrect to infer the IR luminosity from the slope of the rest-frame UV portion of the stellar SED, using the local IRX-$\beta$ relation, at least for this small sub-sample of high-redshift galaxies. We found that the shallow slopes of the attenuation curves for FIR-bright LBGs may be produced, at least in part, by the irregular morphology of stars and dust. However, due to the lack of high-resolution UV and IR data, we cannot reach any quantitative statements. We conclude that a large, statistically significant sample of high-redshift LBGs with high-resolution data is required in order to advance this job further.

%ZZZ IRS end with some paragraph giving the broader context/conclusions from this work?

%
%
%
\section*{Acknowledgements}

M.P.K.\ and K.E.K.C.\ acknowledge support from the UK's Science and Technology Facilities Council (grant numbers ST/M001008/1 and ST/R000905/1). K.E.K.C. is also supported by a Royal Society Leverhulme Trust Senior Research Fellowship (SRF/R1/191013). A.K.\ and M.P.K.\ acknowledge support from the First TEAM grant of the Foundation for Polish Science No. POIR.04.04.00-00-5D21/18-00. A.K.\ also acknowledges support from the Polish National Science Center grants 2016/21/D/ST9/01098 and 2014/15/B/ST9/02111. J.E.G.\ is supported by a Royal Society URF.   I.R.S., U.D., B.G., J.M.S.\ and A.M.S.\ also acknowledge support from STFC (ST/P000541/1). J.L.W. acknowledges support from an STFC Ernest Rutherford Fellowship (ST/P004784/2). M.J.M.~acknowledges the support of the National Science Centre, Poland through the SONATA BIS grant 2018/30/E/ST9/00208. Support for this work was provided by the Polish National Agency for Academic Exchange through the project InterAPS.

The ALMA data used in this paper were obtained under programs ADS/JAO.ALMA\#2012.1.00090.S, \#2015.1.01528.S, \#2016.1.00434.S and \#2017.1.01492.S. ALMA is a partnership of ESO (representing its member states), NSF (USA) and NINS (Japan), together with NRC (Canada) and NSC and ASIAA (Taiwan), in cooperation with the Republic of Chile. The Joint ALMA Observatory is operated by ESO, AUI/NRAO, and NAOJ.

The UKIDSS project is defined in \citet{Lawrence_2007}. Further details on the UDS can be found in Almaini et al. (in prep). UKIDSS uses the UKIRT Wide Field Camera (WFCAM; \citealt{Casali_2007}). The photometric system is described in \citet{Hewett_2006}, and the calibration is described in \citet{Hodgkin_2009}. The pipeline processing and science archive are described in Irwin et al (in prep) and \citet{Hambly_2008}.

This work is based on observations taken by the CANDELS Multi-Cycle Treasury Program with the NASA/ESA HST, which is operated by the Association of Universities for Research in Astronomy, Inc., under NASA contract NAS5-26555.

%
%
%

% \bibliographystyle{mnras}
% \bibliography{refs}

%ZZZ there are a couple of 2016 arxiv papers which must be in press now?

\bsp	% typesetting comment

% \clearpage

\appendix

\section{Tables}

%ZZZ IRS REMOVE Z_PHOT AND RA/DEC and refer reader to ugne/stuart's papers for those numbers
%ZZZ too many sig fig in the SFR columns - quote errors to 1-2 sig fig and values to same number
%
%
%
\begin{landscape}
\begin{table}
\begin{center}
\caption{Target properties. Stellar mass, UV and IR luminosity, followed by UV and IR star formation rate, UV slope, infrared excess, the relative slope of the attenuation curve and the UV bump strength. IDs with stars represent sources with bad rest-frame UV SED fits (see Figure\,\ref{fig:sedsall}).}
\label{tab:prop}
\setlength{\tabcolsep}{1.5 mm}
\begin{normalsize}
\begin{tabular}{lrrrcrcccc}
\hline
\hline
ID & log($M_\ast/{\rm M}_\odot$) & log($L_{\rm UV}/{\rm L}_\odot$) & log($L_{\rm IR}/{\rm L}_\odot$) & ${\rm SFR_{UV}/M_\odot\,yr^{-1}}$ & ${\rm SFR_{IR}/M_\odot\,yr^{-1}}$ & $\beta$ & log(IRX) & $\delta$ & $B$ \\
\hline
AS2UDS.0004.0 & $11.1 \pm 0.2$ & $10.04 \pm 0.06$ & $12.96 \pm 0.04$ & $3 \pm 1$ & $1000 \pm 100\phantom{11}$ & $\phantom{-}1.23 \pm 0.45$ & $0.47 \pm 0.01$ & $          -0.25 \pm 0.10$ & 3.0 \\
AS2UDS.0006.0 & $10.8 \pm 0.2$ & $10.12 \pm 0.17$ & $12.98 \pm 0.04$ & $3 \pm 2$ & $1040 \pm 100\phantom{11}$ & $          -1.21 \pm 0.25$ & $0.46 \pm 0.03$ & $\phantom{-}0.45 \pm 0.13$ & 0.0 \\
AS2UDS.0008.0 & $10.8 \pm 0.2$ & $ 9.95 \pm 0.14$ & $12.75 \pm 0.10$ & $2 \pm 1$ & $610 \pm 150\phantom{11}$ & $          -0.66 \pm 0.21$ & $0.45 \pm 0.03$ & $\phantom{-}0.32 \pm 0.16$ & 0.0 \\
AS2UDS.0014.0 & $10.9 \pm 0.2$ & $10.51 \pm 0.18$ & $13.05 \pm 0.03$ & $8 \pm 4$ & $1210 \pm 70\phantom{111}$ & $          -0.29 \pm 0.39$ & $0.40 \pm 0.03$ & $\phantom{-}0.13 \pm 0.12$ & 1.0 \\
AS2UDS.0016.0 & $11.0 \pm 0.2$ & $10.40 \pm 0.09$ & $12.89 \pm 0.04$ & $6 \pm 1$ & $850 \pm 90\phantom{111}$ & $          -1.00 \pm 0.39$ & $0.40 \pm 0.02$ & $\phantom{-}0.38 \pm 0.07$ & 0.0 \\
AS2UDS.0026.0 & $11.6 \pm 0.6$ & $ 9.90 \pm 0.40$ & $12.78 \pm 0.16$ & $2 \pm 3$ & $660 \pm 180\phantom{11}$ & $          -1.89 \pm 1.27$ & $0.46 \pm 0.06$ & $\phantom{-}0.57 \pm 0.11$ & 0.0 \\
AS2UDS.0039.0 & $11.0 \pm 0.1$ & $ 9.86 \pm 0.10$ & $12.93 \pm 0.03$ & $2 \pm 1$ & $920 \pm 60\phantom{111}$ & $          -0.26 \pm 0.37$ & $0.49 \pm 0.01$ & $\phantom{-}0.26 \pm 0.12$ & 0.0 \\
AS2UDS.0044.0 & $11.2 \pm 0.1$ & $ 9.78 \pm 0.18$ & $12.74 \pm 0.12$ & $1 \pm 1$ & $600 \pm 160\phantom{11}$ & $\phantom{-}0.19 \pm 0.79$ & $0.47 \pm 0.03$ & $\phantom{-}0.00 \pm 0.09$ & 1.0 \\
AS2UDS.0049.0 & $11.0 \pm 0.1$ & $ 9.60 \pm 0.18$ & $12.67 \pm 0.03$ & $1 \pm 1$ & $510 \pm 30\phantom{111}$ & $\phantom{-}0.55 \pm 0.66$ & $0.49 \pm 0.03$ & $\phantom{-}0.07 \pm 0.10$ & 1.0 \\
AS2UDS.0076.0 & $10.8 \pm 0.1$ & $10.71 \pm 0.06$ & $12.79 \pm 0.05$ & $12 \pm 2\phantom{1}$ & $670 \pm 80\phantom{111}$ & $          -1.11 \pm 0.22$ & $0.32 \pm 0.02$ & $\phantom{-}0.32 \pm 0.10$ & 0.0 \\
AS2UDS.0082.0 & $11.1 \pm 0.1$ & $ 9.93 \pm 0.06$ & $12.73 \pm 0.02$ & $2 \pm 1$ & $590 \pm 30\phantom{111}$ & $          -0.31 \pm 0.40$ & $0.45 \pm 0.01$ & $\phantom{-}0.20 \pm 0.08$ & 0.0 \\
AS2UDS.0086.0$^*$ & $ 9.5 \pm 0.1$ & $10.04 \pm 0.12$ & $12.54 \pm 0.16$ & $3 \pm 1$ & $380 \pm 140\phantom{11}$ & $\phantom{-}0.90 \pm 0.17$ & $0.40 \pm 0.03$ & $          -0.31 \pm 0.08$ & 0.0 \\
AS2UDS.0099.0 & $11.3 \pm 0.2$ & $ 9.92 \pm 0.11$ & $12.78 \pm 0.06$ & $2 \pm 1$ & $650 \pm 90\phantom{111}$ & $          -0.13 \pm 0.80$ & $0.46 \pm 0.02$ & $\phantom{-}0.07 \pm 0.09$ & 1.0 \\
AS2UDS.0114.0 & $10.8 \pm 0.2$ & $10.68 \pm 0.09$ & $12.70 \pm 0.25$ & $12 \pm 2\phantom{1}$ & $540 \pm 510\phantom{11}$ & $          -1.05 \pm 0.20$ & $0.30 \pm 0.05$ & $\phantom{-}0.20 \pm 0.09$ & 0.0 \\
AS2UDS.0121.0 & $10.6 \pm 0.1$ & $10.35 \pm 0.18$ & $12.90 \pm 0.09$ & $5 \pm 3$ & $870 \pm 190\phantom{11}$ & $          -1.16 \pm 0.22$ & $0.41 \pm 0.03$ & $\phantom{-}0.38 \pm 0.16$ & 0.0 \\
AS2UDS.0121.1 & $10.4 \pm 0.2$ & $10.41 \pm 0.23$ & $12.97 \pm 0.04$ & $6 \pm 4$ & $1020 \pm 90\phantom{111}$ & $          -1.63 \pm 0.43$ & $0.41 \pm 0.04$ & $\phantom{-}0.38 \pm 0.14$ & 0.0 \\
AS2UDS.0156.0 & $11.3 \pm 0.1$ & $10.26 \pm 0.08$ & $12.69 \pm 0.08$ & $4 \pm 1$ & $530 \pm 90\phantom{111}$ & $\phantom{-}0.22 \pm 0.17$ & $0.39 \pm 0.02$ & $          -0.18 \pm 0.14$ & 0.0 \\
AS2UDS.0162.1 & $10.0 \pm 0.1$ & $ 9.70 \pm 0.10$ & $12.58 \pm 0.11$ & $1 \pm 1$ & $410 \pm 90\phantom{111}$ & $          -0.44 \pm 0.29$ & $0.46 \pm 0.02$ & $\phantom{-}0.07 \pm 0.07$ & 0.0 \\
AS2UDS.0209.0 & $10.8 \pm 0.1$ & $10.43 \pm 0.05$ & $12.81 \pm 0.17$ & $6 \pm 1$ & $700 \pm 230\phantom{11}$ & $          -0.47 \pm 0.21$ & $0.38 \pm 0.03$ & $\phantom{-}0.07 \pm 0.09$ & 1.0 \\
AS2UDS.0223.0 & $10.6 \pm 0.2$ & $10.02 \pm 0.15$ & $12.59 \pm 0.16$ & $2 \pm 1$ & $420 \pm 170\phantom{11}$ & $          -0.55 \pm 0.44$ & $0.41 \pm 0.04$ & $\phantom{-}0.07 \pm 0.10$ & 0.0 \\
AS2UDS.0228.0$^*$ & $10.1 \pm 0.5$ & $ 9.98 \pm 0.22$ & $12.32 \pm 0.12$ & $2 \pm 1$ & $230 \pm 60\phantom{111}$ & $          -0.98 \pm 0.32$ & $0.37 \pm 0.04$ & $\phantom{-}0.20 \pm 0.06$ & 0.0 \\
AS2UDS.0266.0 & $10.6 \pm 0.2$ & $ 9.68 \pm 0.48$ & $12.50 \pm 0.16$ & $1 \pm 2$ & $350 \pm 140\phantom{11}$ & $\phantom{-}0.32 \pm 0.70$ & $0.45 \pm 0.07$ & $\phantom{-}0.00 \pm 0.11$ & 1.0 \\
AS2UDS.0268.0 & $11.2 \pm 0.2$ & $10.24 \pm 0.06$ & $12.75 \pm 0.15$ & $4 \pm 1$ & $610 \pm 290\phantom{11}$ & $\phantom{-}0.46 \pm 0.20$ & $0.40 \pm 0.03$ & $          -0.37 \pm 0.13$ & 1.0 \\
AS2UDS.0278.0 & $10.7 \pm 0.1$ & $10.33 \pm 0.03$ & $12.43 \pm 0.08$ & $5 \pm 1$ & $300 \pm 50\phantom{111}$ & $          -0.88 \pm 0.11$ & $0.32 \pm 0.02$ & $\phantom{-}0.32 \pm 0.15$ & 0.0 \\
AS2UDS.0291.0 & $10.5 \pm 0.1$ & $10.61 \pm 0.05$ & $12.60 \pm 0.11$ & $10 \pm 1\phantom{1}$ & $430 \pm 120\phantom{11}$ & $          -0.69 \pm 0.24$ & $0.30 \pm 0.03$ & $\phantom{-}0.07 \pm 0.09$ & 0.0 \\
AS2UDS.0300.0 & $11.1 \pm 0.1$ & $10.17 \pm 0.12$ & $12.66 \pm 0.25$ & $4 \pm 1$ & $500 \pm 310\phantom{11}$ & $          -1.31 \pm 0.34$ & $0.40 \pm 0.05$ & $\phantom{-}0.45 \pm 0.10$ & 1.0 \\
AS2UDS.0308.0 & $10.8 \pm 0.1$ & $10.50 \pm 0.21$ & $12.80 \pm 0.12$ & $8 \pm 5$ & $690 \pm 170\phantom{11}$ & $          -0.74 \pm 0.30$ & $0.36 \pm 0.04$ & $\phantom{-}0.26 \pm 0.16$ & 0.0 \\
AS2UDS.0403.0 & $10.4 \pm 0.2$ & $ 9.88 \pm 0.10$ & $12.77 \pm 0.20$ & $2 \pm 1$ & $640 \pm 280\phantom{11}$ & $          -0.89 \pm 0.42$ & $0.46 \pm 0.03$ & $\phantom{-}0.32 \pm 0.13$ & 0.0 \\
AS2UDS.0436.0 & $10.3 \pm 0.3$ & $ 9.63 \pm 0.18$ & $12.50 \pm 0.10$ & $1 \pm 1$ & $340 \pm 80\phantom{111}$ & $          -0.68 \pm 0.36$ & $0.46 \pm 0.03$ & $\phantom{-}0.13 \pm 0.13$ & 0.0 \\
AS2UDS.0441.0 & $10.7 \pm 0.2$ & $ 9.92 \pm 0.13$ & $12.68 \pm 0.20$ & $2 \pm 1$ & $520 \pm 330\phantom{11}$ & $\phantom{-}0.43 \pm 0.46$ & $0.44 \pm 0.04$ & $          -0.06 \pm 0.12$ & 1.0 \\
AS2UDS.0451.0 & $11.1 \pm 0.1$ & $10.20 \pm 0.13$ & $12.71 \pm 0.10$ & $4 \pm 1$ & $560 \pm 130\phantom{11}$ & $          -1.16 \pm 0.27$ & $0.40 \pm 0.03$ & $\phantom{-}0.51 \pm 0.18$ & 1.0 \\
AS2UDS.0461.0 & $10.4 \pm 0.1$ & $ 9.75 \pm 0.18$ & $12.71 \pm 0.09$ & $1 \pm 1$ & $560 \pm 140\phantom{11}$ & $\phantom{-}0.06 \pm 0.33$ & $0.47 \pm 0.03$ & $\phantom{-}0.13 \pm 0.10$ & 1.0 \\
AS2UDS.0480.0 & $ 9.6 \pm 0.2$ & $ 9.68 \pm 0.33$ & $12.48 \pm 0.08$ & $1 \pm 1$ & $330 \pm 60\phantom{111}$ & $          -1.23 \pm 0.91$ & $0.45 \pm 0.05$ & $\phantom{-}0.32 \pm 0.05$ & 0.0 \\
AS2UDS.0481.1$^*$ & $10.2 \pm 0.1$ & $ 9.69 \pm 0.23$ & $12.74 \pm 0.12$ & $1 \pm 1$ & $600 \pm 160\phantom{11}$ & $          -1.10 \pm 0.62$ & $0.49 \pm 0.04$ & $\phantom{-}0.38 \pm 0.11$ & 0.0 \\
AS2UDS.0485.0 & $11.1 \pm 0.1$ & $10.22 \pm 0.07$ & $12.44 \pm 0.07$ & $4 \pm 1$ & $300 \pm 50\phantom{111}$ & $          -0.28 \pm 0.16$ & $0.35 \pm 0.02$ & $\phantom{-}0.13 \pm 0.15$ & 0.0 \\
AS2UDS.0514.0 & $11.0 \pm 0.1$ & $ 9.97 \pm 0.09$ & $12.68 \pm 0.04$ & $2 \pm 1$ & $520 \pm 50\phantom{111}$ & $\phantom{-}0.66 \pm 0.51$ & $0.43 \pm 0.02$ & $          -0.06 \pm 0.07$ & 0.0 \\
AS2UDS.0573.0 & $11.3 \pm 0.2$ & $11.39 \pm 0.08$ & $12.65 \pm 0.10$ & $59 \pm 12$ & $490 \pm 110\phantom{11}$ & $          -1.22 \pm 0.14$ & $0.10 \pm 0.04$ & $\phantom{-}0.07 \pm 0.09$ & 0.0 \\
AS2UDS.0583.0 & $10.6 \pm 0.2$ & $10.10 \pm 0.19$ & $12.26 \pm 0.06$ & $3 \pm 2$ & $200 \pm 30\phantom{111}$ & $          -0.64 \pm 0.25$ & $0.33 \pm 0.04$ & $\phantom{-}0.00 \pm 0.20$ & 0.0 \\
AS2UDS.0596.0 & $10.6 \pm 0.2$ & $10.23 \pm 0.20$ & $12.84 \pm 0.09$ & $4 \pm 2$ & $750 \pm 170\phantom{11}$ & $          -0.93 \pm 0.56$ & $0.42 \pm 0.04$ & $\phantom{-}0.32 \pm 0.14$ & 0.0 \\
AS2UDS.0678.0 & $10.5 \pm 0.3$ & $ 9.89 \pm 0.20$ & $12.54 \pm 0.06$ & $2 \pm 1$ & $370 \pm 50\phantom{111}$ & $          -1.25 \pm 0.50$ & $0.42 \pm 0.03$ & $\phantom{-}0.38 \pm 0.12$ & 0.0 \\
AS2UDS.0712.0 & $11.0 \pm 0.1$ & $10.30 \pm 0.14$ & $13.01 \pm 0.03$ & $5 \pm 2$ & $1100 \pm 80\phantom{111}$ & $          -0.66 \pm 0.34$ & $0.43 \pm 0.02$ & $\phantom{-}0.20 \pm 0.13$ & 0.0 \\
\hline
\end{tabular}
\end{normalsize}
\end{center}
\end{table}
\end{landscape}

\label{lastpage}
\end{document}